\input harvmac
\input amssym.def
\input amssym.tex
\parskip=4pt \baselineskip=12pt
\hfuzz=20pt
\parindent 10pt

\def\newsubsec#1{\global\advance\subsecno by1\message{(\secsym\the\subsecno.
#1)} \ifnum\lastpenalty>9000\else\bigbreak\fi
\noindent{\bf\secsym\the\subsecno. #1}\writetoca{\string\quad
{\secsym\the\subsecno.} {#1}}}%

\global\newcount\subsubsecno \global\subsubsecno=0
\def\subsubsec#1{\global\advance\subsubsecno
by1\message{(\secsym\the\subsecno.\the\subsubsecno. #1)}
\ifnum\lastpenalty>9000\else\bigbreak\fi
\noindent{\bf\secsym\the\subsecno.\the\subsubsecno. #1}\writetoca{\string\quad
{\secsym\the\subsecno.\the\subsubsecno.} {#1}}\par\nobreak\medskip\nobreak}

\def\newsubsubsec#1{\global\advance\subsubsecno
by1\message{(\secsym\the\subsecno.\the\subsubsecno.
#1)} \ifnum\lastpenalty>9000\else\bigbreak\fi
\noindent{\bf\secsym\the\subsecno.\the\subsubsecno. #1}\writetoca{\string\quad
{\secsym\the\subsecno.\the\subsubsecno.} {#1}}}%

\def\nt{\noindent}
\def\nl{\hfill\break}

\def\np{\vfill\eject}
\def\nd{\vfill\eject\end}

\def\bu{$\bullet$}

\def\mt{\mapsto}
\font\tfont=cmbx12 scaled\magstep1 
\font\male=cmr9

\def\Box{
\vbox{
\halign to5pt{\strut##&
\hfil ## \hfil \cr
&$\kern -0.5pt
\sqcap$ \cr
\noalign{\kern -5pt
\hrule}
}}~}

\input epsf.tex
\newcount\figno
\figno=0
\def\fig#1#2#3{
\par\begingroup\parindent=0pt\leftskip=1cm\rightskip=1cm\parindent=0pt
\baselineskip=11pt
\global\advance\figno by 1
\midinsert
\epsfxsize=#3
\centerline{\epsfbox{#2}}
\vskip 12pt
#1\par
\endinsert\endgroup\par}
\def\figlabel#1{\xdef#1{\the\figno}}
\def\encadremath#1{\vbox{\hrule\hbox{\vrule\kern8pt\vbox{\kern8pt
\hbox{$\displaystyle #1$}\kern8pt}
\kern8pt\vrule}\hrule}}

\def\llra#1{\mathop{\longrightarrow}\limits_{#1}}
\def\down{\raise1.5pt\hbox{$\phantom{a}_2$}\downarrow}

\def\downa{\raise1.5pt\hbox{$\phantom{a}_{2\atop m_2}$}\downarrow}
\def\upa{\phantom{a}_{4\atop m_2}\uparrow}

\def\hp{\hat{\varphi}}

\def\hx{\hat{X}}

\def\({\left(}
\def\){\right)}
\def\eps{\epsilon}
 
\def\lra{\longrightarrow}

\def\tL{{\Lambda'}}
\def\dia{{$\diamondsuit$}}
  
\def\vf{\varphi}
\def\ha{{\textstyle{1\over2}}}
\def\qrt{{\textstyle{1\over4}}}
\def\trha{{\textstyle{3\over2}}}
\def\fha{{\textstyle{5\over2}}}
\def\iha{{\textstyle{i\over2}}}
\def\pd{\partial}

\def\bac{{C\kern-5.5pt I}}
\def\bbz{Z\!\!\!Z}
\def\bbc{C\kern-6.5pt I}
\def\bbr{I\!\!R}
\def\bbn{I\!\!N}
\def\a{\alpha}
\def\b{\beta}
\def\d{\delta}

\def\vr{\vert}

\def\s{\sigma}
\def\G{\Gamma}
\def\y{\eta}
\def\l{\lambda}

\def\m{\mu}
\def\D{\Delta}

\def\ca{{\cal A}} \def\cb{{\cal B}} \def\cc{{\cal C}}
  \def\cf{{\cal F}}
\def\cg{{\cal G}} \def\ch{{\cal H}} 
 \def\ck{{\cal K}} \def\cl{{\cal L}}
\def\cm{{\cal M}} \def\cn{{\cal N}} 
  
\def\cs{{\cal S}}

\def\L{\Lambda}
\def\r{\rho}
\def\cgc{{\cg^\bac}}



\nref\AOSV{M. Aganagic, H. Ooguri, N. Saulina and C. Vafa, Nucl.
Phys. {\bf B715}, 304 (2005)
  hep-th/0411280.}

\nref\LuTs{D. Lust and D. Tsimpis,    JHEP
0502:027 (2005)  hep-th/0412250.}

\nref\GLMW{J.P. Gauntlett, S. Lee, T. Mateos and D. Waldram, JHEP
0508:030 (2005)  hep-th/0505207.}

\nref\ABMS{X. Arsiwalla, R. Boels, M. Marino and A. Sinkovics,  hep-th/0509002.}

\nref\JaMa{D. Jafferis and J. Marsano,    hep-th/0509004.}

\nref\CCGPSS{
N.  Caporaso, M. Cirafici, L. Griguolo, S. Pasquetti, D. Seminara and R.J. Szabo,
   \& hep-th/0511043.}

\nref\AFT{R. D'Auria, S. Ferrara and M. Trigiante, hep-th/0512248
and Proceedings of   IV International Symposium
'Quantum Theory and Symmetries' (15-21 August 2005, Varna, Bulgaria),
to appear.}

\nref\FeLl{S. Ferrara and  M.A. Lledo, JHEP 0005:008 (2000)
hep-th/0002084.}

\nref\FeSo{S. Ferrara and E. Sokatchev,   JHEP 0005:038 (2000)
hep-th/0003051.}

\nref\BFFMZ{M. Billo, D. Fabbri, P. Fre, P. Merlatti and  A.
Zaffaroni,   Nucl. Phys. {\bf B591}, 139 (2000)  hep-th/0005220.}

\nref\KMP{I.I. Kogan, S. Mouslopoulos and A. Papazoglou,
Phys. Lett. {\bf B503}, 173
(2001)  hep-th/0011138.}

\nref\CGLP{M. Cvetic, G.W. Gibbons, Hong Lu and  C.N. Pope,
Nucl. Phys. {\bf B617}, 151
(2001)  hep-th/0102185.}

\nref\DIK{F. Delduc, E. Ivanov and S. Krivonos,
Phys. Lett. {\bf B529}, 233
(2002)  hep-th/0111106.}

\nref\PoSt{M. Porrati and A. Starinets,
Phys. Lett. {\bf B532}, 48
(2002)  hep-th/0201261.}

\nref\KrRa{F. Kristiansson and P. Rajan,   JHEP
0304:009 (2003) hep-th/0303202.}

\nref\LePe{R.G. Leigh and A.C. Petkou,
JHEP 0306:011 (2003) hep-th/0304217.}

\nref\SaYo{M. Sakaguchi and K. Yoshida,
   Nucl. Phys. {\bf B681}, 137
   (2004) hep-th/0310035.}

\nref\Ruhl{R. Manvelyan and W. Ruhl,
 Phys. Lett. {\bf B593}, 253
 (2004) hep-th/0403241; ~~W. Ruhl, Phys. Lett. {\bf B605}, 413
(2005) hep-th/0409252; ~~R. Manvelyan and W. Ruhl,
 Phys. Lett. {\bf B613}, 197
(2005) hep-th/0412252.}

\nref\SeSu{E. Sezgin and P. Sundell,  hep-th/0511296
and Proceedings of   IV International Symposium
'Quantum Theory and Symmetries' (15-21 August 2005, Varna, Bulgaria),
to appear.}

\nref\Dirac{P.A.M. Dirac, J. Math. Phys. {\bf 4}, 901 (1963).}

\nref\Fro{C. Fronsdal, Rev. Mod. Phys. {\bf 37}, 221 (1965).}

\nref\Evans{N.T. Evans, J. Math. Phys. {\bf 8}, 170 (1967).}

\nref\Froa{C. Fronsdal,Phys. Rev. {\bf D10}, 589 (1974).}

\nref\Frob{C. Fronsdal,  Phys. Rev. {\bf D12}, 3819 (1975).}

\nref\FFa{M. Flato and C. Fronsdal, Lett. Math. Phys. {\bf 2}, 421 (1978).}

\nref\FFb{M. Flato and C. Fronsdal,  Phys. Lett. {\bf 97B}, 236 (1980).}

\nref\FFc{M. Flato and C. Fronsdal,  J. Math. Phys. {\bf 22}, 1100 (1981).}

\nref\Froc{C. Fronsdal, Phys. Rev. {\bf D26}, 1988 (1982).}


\lref\FD{F.A. Dolan, ``Character formulae and partition functions
in higher dimensional conformal field theory'', hep-th/0508031 (2005).}

\lref\Cala{F.~Calogero, J. Math. Physics {\bf 12},  419 (1971).}

\lref\Sut{B.~Sutherland,  Phys. Rev. {\bf A5},  1372  (1972).}

\lref\Moser{J.~Moser,   Adv. Math. {\bf 16},  197-220 (1975);
~in: {\it Dynamical Systems, Theory and Applications},
Lecture Notes in Physics {\bf 38}
(Springer-Verlag, 1975).}

\lref\CMR{F.~Calogero, C.~Marchioro  and O.~Ragnisco,  Lett. Nuovo
Cim. {\bf 13},  383 (1975).}

\lref\Calb{F.~Calogero, Lett. Nuovo Cim. {\bf 13},  411 (1975).}

\lref\Corr{E. Corrigan,  math-ph/0411043.}

\lref\SaTa{R. Sasaki and K. Takasaki,   hep-th/0510035.}

\lref\OPa{M.A.~Olshanetsky and A.M.~Perelomov,
  Inv. Math. {\bf 37}, 93 (1976).}

\lref\OPb{M.A.~Olshanetsky and A.M.~Perelomov,
Phys. Rep. {\bf 71},  314 (1981).}

\lref\ByFr{A.G. Bytsko and  A. Fring,  Phys. Lett. {\bf B454}, 59 (1999)
hep-th/9812115.}

\lref\KS{A.W. Knapp and E.M. Stein, Ann.
Math. {\bf 93} (1971) 489; ~Inv. Math. {\bf 60}, 9 (1980).}
\lref\Koller{K. Koller, Comm. Math. Phys. {\bf 40}, 15 (1975).}

\lref\FGGP{S. Ferrara, R. Gatto, A.F. Grillo and G. Parisi,
Lett. Nuovo Cim. {\bf 4}, 115 (1972).}

\lref\Nic{D.Z. Freedman and H. Nicolai,
Nucl. Phys. {\bf B237}, 342 (1984);
~~H. Nicolai and E. Sezgin,
Phys. Lett. {\bf B143}, 389 (1984).}

\lref\Dix{J. Dixmier, {\it Enveloping Algebras} (North
Holland, New York, 1977).}

\lref\Helg{S. Helgason, {\it Differential Geometry, Lie Groups
and Symmetric Spaces}, (Academic Press, New York, 1978).}

\lref\Kac{V.G. Kac, {\it Infinite Dimensional Lie Algebras}, 3rd ed.
(Cambridge University Press, 1990).}

\lref\BGG{N.N. Bernstein, I.M. Gelfand and S.I. Gelfand,
Funkts. Anal. Prilozh. {\bf 5}, 1 (1971); English translation, Funct. Anal. Appl.
{\bf 5}, 1 (1971).}

\lref\GeSh{I.M. Gelfand and G.E. Shilov,
Generalised Functions, vol. 1, (Academic Press, New York, 1964).}

\lref\DoSe{V.K. Dobrev and E. Sezgin,
in: Lecture Notes in Physics,
vol. 379 (Springer-Verlag, Berlin, 1990) p. 227.}

\lref\DoSekm{V.K. Dobrev and E. Sezgin,
Int. J. Mod. Phys. {\bf A6} (1991) 4699.}

\lref\DoMo{V.K. Dobrev and P.J. Moylan,
Phys. Lett. {\bf 315B}, 292 (1993).}

\lref\DoMop{V.K. Dobrev and P.J. Moylan,
in preparation.}

\lref\DMPPT{V.K. Dobrev, G. Mack, V.B. Petkova, S.G. Petrova and I.T.
Todorov, {\it Harmonic Analysis on the $n$ - Dimensional
Lorentz Group and Its Applications to Conformal Quantum Field
Theory}, Lecture Notes in Physics, No 63, 280 pages (Springer
Verlag, Berlin-Heidelberg-New York, 1977).}

\lref\Domult{V.K. Dobrev, Lett. Math. Phys. {\bf 9}, 205 (1985);
~~Talk at the Conference on Algebraic Geometry and Integrable
Systems (Oberwolfach, 1984) ~and ~~ICTP, Trieste, preprint IC/85/9 (1985).}

\lref\Dosv{V.K. Dobrev, Lett. Math. Phys. {\bf 22}, 251 (1991).}

\lref\Dojmp{V.K. Dobrev, J. Math. Phys. {\bf 26}, 235 (1985).}

\lref\Doqg{V.K. Dobrev, J. Math. Phys. {\bf 33}, 3419 (1992);
Phys. Lett. {\bf B341}, 133 (1994).}

\lref\DoPe{V.K. Dobrev and V.B. Petkova,
Rep. Math. Phys. {\bf 13}, 233 (1978).}

\lref\Doa{V.K. Dobrev,
Rep. Math. Phys. {\bf 25}, 159-181 (1988).}

\lref\Doch{V.K. Dobrev,
Phys. Lett. {\bf 186B}, 43 (1987);
~~Suppl. Rend. Circolo Mat. Palermo, Serie II, Numero 14
(1987) 25;\nl V.K. Dobrev and A.Ch. Ganchev,
Mod. Phys. Lett. {\bf A3}, 127 (1988).}

\lref\KS{A.W. Knapp and E.M. Stein,
Ann. Math. {\bf 93}, 489-578 (1971); ~II:
Inv. Math. {\bf 60}, 9-84 (1980).}

\lref\KZ{A.W. Knapp and G.J. Zuckerman,   .
in: Lecture Notes in
Math., vol. 587, pp. 138-159 (Berlin, Springer-Verlag, 1977);
Ann. Math. {\bf 116}, 389-501 (1982).}

\lref\La{R.P. Langlands, "On the classification of irreducible
representations of real algebraic groups", preprint,
 Institute for Advanced Study, Princeton (1973);
published in: {\it Representation Theory and Harmonic Analysis
on Semisimple Lie Groups},
eds. P. Sally and D. Vogan, Math. Surveys and Monographs vol. 31,
(AMS, 1989) pp. 101-170.}

\lref\FFP{ J.F. Nunez, W.G. Fuertes and A.M. Perelomov,
math-ph/0406067.}


\centerline{{\tfont Invariant Differential Operators and }}
\vskip 2truemm
\centerline{{\tfont Characters  of the AdS$_4$ Algebra}}

\vskip 1.5cm

\centerline{{\bf V.K. Dobrev}}
\vskip 0.5cm

\centerline{Institute of Nuclear Research and Nuclear Energy}
\centerline{Bulgarian Academy of Sciences}
\centerline{72 Tsarigradsko Chaussee, 1784 Sofia, Bulgaria}
\vskip 1.5cm

\centerline{{\bf Abstract}}
\midinsert\narrower{\male
The aim of this paper is to apply systematically to AdS$_4$   some modern tools in the representation
theory of Lie algebras which are easily generalised to the supersymmetric and
quantum group settings and necessary for applications to string theory and
integrable models. Here we introduce the necessary representations of
the AdS$_4$ algebra and group.  We give explicitly
all singular (null) vectors of the reducible AdS$_4$ Verma modules.
These are used  to obtain the AdS$_4$ invariant differential operators.
Using this we display a new structure - a diagram involving four
partially equivalent reducible representations
one of which contains all finite-dimensional irreps of the AdS$_4$ algebra.
We study in more detail   the  cases involving UIRs, in particular, the
Di and the Rac singletons, and the massless UIRs.
In the massless case we discover the structure
of sets of ~$2s_0-1$~ conserved currents for each spin~$s_0$~ UIR, $s_0=1,\trha,\ldots$
~All massless cases are contained in a one-parameter subfamily of the quartet diagrams
mentioned above, the parameter being the spin $s_0$.
Further  we give the classification of the  ~$so(5,\bbc)$~ irreps
 presented in a diagrammatic way which makes easy the derivation
of all character formulae.
The paper concludes with a speculation on the possible applications
of the character formulae to integrable models.
 }\endinsert

\noindent
MSC: 17B10, 22E47, 81R05\nl
PACS: 02.20.Qs, 02.20.Sv, 11.25.Hf

\vskip 1.5cm

\newsec{Introduction}

\nt
The AdS/CFT correspondence
between gravitational and gauge forces in general states that a
bulk supergravity theory resp. closed string theory is equivalent
to a Yang-Mills resp. open string theory on the boundary of
space-time (holography). This correspondence is best understood
for four-dimensional  N=4 supersymmetric Yang-Mills, which is dual to
supergravity on AdS5, and which can be explained by D3-branes in
type IIB superstring theory. It is also important that string theories have
conformal  symmetry and thus may be viewed as
ultraviolet limits of integrable (but not conformally
invariant) models; or the latter may be viewed as
deformations, including quantum group deformations,
of conformal models. However, much work remains to be done
in order to establish  analogous results  for dimensions $D \neq 4$
and also  for deformations of the original anti de
Sitter backgrounds following recent developments in
\refs{\AOSV--\AFT}.
In particular, for ~$D=3$~ we need to
develop the representation theory of the ~$AdS_4$~ algebra ~$so(3,2)$~
in a formalism that will be suitable for the purposes of the AdS/CFT
correspondence, supersymmetrizations and $q$-deformations along the lines of
\refs{\FeLl--\SeSu}.

The AdS$_4$ algebra ~$so(3,2)$~ attracted attention very early
since it has truly remarkable unitary
representations known as singletons, which were first discovered by Dirac in
1963 \Dirac. These representations have been extensively studied by Fronsdal,
 Flato and  Evans \refs{\Fro--\Froc}. There are two singleton
representations, called Di
and Rac. In terms of the lowest energy value ~$E_0$~ and the spin $s_0\,$, the
Di has ~$E_0 = 1,\ s_0 = 1/2$~ while the Rac has ~$E_0 = 1/2,\ s_0
= 0$. These representations have remarkably reduced spectrum (weight
spaces). Consequently, the singleton field theory has a very large gauge
symmetry which enables one to gauge away the singleton fields everywhere except
on the boundary of the anti de Sitter space \FFc. Moreover, the direct product
of two singletons decomposes into infinitely many massless states of the anti
de Sitter group \FFa.

However, the above results and the methods applied are not
suitable for the envisaged generalizations. Thus, the aim of this
paper is to apply systematically some modern tools in the
representation theory of Lie algebras \Doa,\Dojmp,\Domult{} which
are easily generalised to the supersymmetric and quantum group
settings \Doqg. These generalizations will be done in sequel(s) of
the present paper taking into account also  work done for the
AdS$_4$ algebra in \DoSe{}, its superpartners \Nic, its Kac-Moody
counterpart \DoSekm, and its quantum group deformation \DoMo.
For $so(d,2)$, $d>4$, similar technique was applied in \FD.

The paper is organized as follows.
In Section  2 we give preliminaries including facts about the
~$so(3,2)$~ algebra and the group ~$SO_0(3,2)$.
In Section  3 we   adapt the approach of \Doa{} to the current setting
introducing the elementary representations (ERs) of ~$SO_0(3,2)$, relating them
to the Verma modules over the complexification $so(5,\bbc)$ of $so(3,2)$.
Then we give the singular (null) vectors of the reducible Verma modules
and use them (a la \Doa) to obtain the invariant differential operators
between the reducible ERs. In Section  4 we apply these differential operators
to the cases involving UIRs. For the massless case we discover the structure
of sets of ~$2s_0-1$~ conserved currents for each spin~$s_0$~ UIR, $s_0=1,\trha,\ldots$
~Another structure parametrized by the same values of $s_0$ is a
quartet diagram of partially equivalent reducible ERs involving the massless case for that $s_0\,$,
a massive UIR (of spin $s_0-1$) and an  ER containing finite-dimensional
irrep of dimension $s_0(4s_0^2-1)/3$.
In Section  5 we give the classification of the  ~$so(5,\bbc)$~ irreps.
This is presented in a diagrammatic way which makes easy the derivation
of the character formulae of these irreps which is done in Sec. 6.
Section 6 is concluded with a speculation on the possible applications
of the derived character formulae to integrable models.

\np

\newsec{Preliminaries}

\newsubsec{Lie algebra.}~~~
We start with the complexification ~$\cg^\bac ~=~ so(5,\bbc) ~=~ B_2$~
of the algebra ~$\cg ~=~ so(3,2)$.
We use the standard definition of ~$\cg^\bac$~ given in terms of
the Chevalley generators $X^\pm_i ~, ~H_i ~, ~i=1,2$,  by the relations~:
\eqna\com
$$\eqalignno{ &[H_j~, ~H_k] ~ = ~ 0 ~, ~~~[H_j~, ~X^\pm_k] ~ = ~ \pm
a_{jk} X^\pm_k ~,
~~~[X^+_j ~, ~X^-_k] ~ = ~
\d_{jk} ~H_j ~, &\com a\cr
&\sum_{m=0}^n ~(-1)^m ~\left({n \atop m}\right)
~\left(X^\pm_j\right)^m
~X^\pm_k ~\left(X^\pm_j\right)^{n-m} ~=~ 0 ~, ~~j \neq k ~, ~~n = 1 -
a_{jk} ~,
&\com b\cr }$$
where
\eqn\cart{ (a_{jk}) ~=~ (\a^\vee_j , \a_k) ~=~
\pmatrix{2 & -2\cr -1 &2} }
is the Cartan matrix of $\cg^\bac$,
~~$\a^\vee_j ~\equiv~ {2 \a_j\over (\a_j , \a_j)}$~ is the co-root of $\a_j\,$,
~~ $(\cdot , \cdot)$ is the scalar product of the roots, so that the
non-zero products between the simple roots are:
$(\a_1,\a_1) = 2$, $(\a_2,\a_2) = 4$, $(\a_1,\a_2) = -2$.
The elements $H_i$ span the Cartan subalgebra $\ch$ of $\cg^\bac$,
while the elements $X^\pm_i$ generate the subalgebras $\cg^\pm$.
We shall use the standard triangular decomposition
\eqn\deca{\cg^\bac ~=~ \cg_+\otimes \ch\oplus \cg_- ~, ~~\cg_\pm ~\equiv
~\mathop{\oplus}\limits_{\a\in\D^\pm} ~\cg_\a ~, }
where $\D^+$, $\D^-$, are the sets of positive, negative, roots, resp.,
and ~$\dim~\cg_\a ~=~ 1$. Explicitly    we have:
\eqn\roots{
 \D^\pm =~ \{ \pm\a_1 ~, ~\pm\a_2 ~, ~\pm\a_3 ~, ~\pm\a_4 \}
 \ , \qquad \a_3 = \a_1+\a_2\,, ~~~\a_4 = 2\a_1 + \a_2 \ . }

Let us denote the root space vector of ~$\cg_\a$~ by ~$X_\a$, or more explicitly:
~$X^\pm_k ~\equiv~ X_{\pm\a_k}\,$, ~$k=1,2,3,4$. To give the full
Cartan-Weyl basis we need to define also $X_k^\pm\,$, ~$k=3,4$, for which we follow \Dosv:
\eqn\cww{X^\pm_3 = \pm [X^\pm_1, X^\pm_2], \qquad X^\pm_4 = \pm \ha\, [  X^\pm_1, X^\pm_3]
 ~. }
  Then we have:
\eqn\crr{ H_3 ~\equiv ~[ X^+_3 , X^-_3 ]  ~=~ H_1 + 2H_2 ~, ~~~
H_4 ~\equiv~ [ X^+_4 , X^-_4 ]  ~=~ H_1 + H_2 ~. }

Note that for ~$H_k$~ also holds:
\eqn\cons{ \l (H_k) ~=~ (\l,\a^\vee_k) \ , \qquad \forall\,\l\in\ch^* \ , \quad k=1,2,3,4 .}

The algebra ~$\cg=so(3,2)$~ is a maximally split real form \Helg{} of $\cgc$ so we can use
the same basis (but over $\bbr$) and the same root system. Thus, we can use the second order
Casimir of ~$\cgc$~:
\eqn\casi{\eqalign{
\cc_2 ~=&~ \ha (X^+_1  X^-_1 + X^-_1  X^+_1) \ +\
X^+_2  X^-_2 + X^-_2  X^+_2 \ +\
\ha (X^+_3  X^-_3 + X^-_3  X^+_3)
\ +\cr &+~ X^+_4  X^-_4 + X^-_4 X^+_4 \ +\
\ha H_1^2 \ +\  H_2^2 \ +\  H_1 H_2 \ .}}
It useful to relate the Cartan-Weyl basis given above with the   ~$so(3,2)$~
generators ~$X_{AB}$~:
\eqna\crab
$$\eqalignno{ & H_1 ~=~ 2iX_{12}  \ ,
~~~ X^\pm_1 ~=~  X_{10} \pm i X_{20}  \ , &\crab a\cr
&H_2 ~=~  X_{34} - iX_{12}\ ,  ~~~ X^\pm_2 ~=~ \mp \ha \left(X_{23} \pm X_{24} + iX_{14} \pm
iX_{13}  \right)  \ , &\crab b\cr
& X^\pm_3 ~=~ \mp i \( X_{03} \pm X_{04} \) \ ,   &\crab c\cr
& X^\pm_4 ~=~ \ha \left( X_{24} \pm X_{23} \mp iX_{14}
-iX_{13}  \right) \ .  &\crab d\cr
 } $$
It is easy to see that the ten generators $X_{AB} = - X_{BA}$, $A,B =
0,1,2,3,4,$  satisfy the standard $so(3,2)$ commutation relations:\foot{Note that
often are used the generators ~$M_{AB} ~=~ iX_{AB}\,$.}
\eqn\sodef{[X_{AB},X_{CD}] ~=~
\y_{AC}X_{BD} + \y_{BD}X_{AC} - \y_{AD}X_{BC} - \y_{BC}X_{AD}
 ~. }
where ~$\y_{11}=\y_{22}=\y_{33}=-\y_{00}=-\y_{44}=1$, $\y_{jk} =0$ if $j\neq k$.

\newsubsec{Finite-dimensional realization.}~~~
It is useful to have a finite-dimensional realization of ~$\cg$~ which we take from \Dojmp{}
(cf. (2.18) which we restrict from $so(4,2)$ to $so(3,2)$):\foot{It is different
from the one in \Nic{} which we used in \DoSe.}
\eqna\fd
$$\eqalignno{ &  X_{12} ~=~ -\iha \pmatrix{ \s_3 & 0 \cr 0 & \s_3}\ ,
 \qquad X_{a3} ~=~ \iha \pmatrix{ 0 & \s_a  \cr  \s_a & 0} \ ,  ~ a=1,2, \cr
& X_{04} ~=~ \iha \pmatrix{ 0 & 1_2 \cr 1_2 & 0} \ ;
 &\fd a\cr
&X_{34} ~=~ \ha \pmatrix{1_2  & 0 \cr 0 & -1_2 }\ ,
\qquad X_{a0} ~=~ \ha \pmatrix{  \s_a & 0  \cr 0 & - \s_a } \ , ~ a=1,2, &\fd b\cr
& X_{03} ~=~ \iha \pmatrix{ 0 & 1_2 \cr -1_2 & 0} \ ,
\qquad X_{a4} ~=~ \iha \pmatrix{ 0 & \s_a  \cr  -\s_a & 0} \ , ~ a=1,2.
}$$
The hermiticity properties of this defining realisation are:
\eqna\herm
$$\eqalignno{
&X^{\dag}_{AB} ~=~ -X_{AB}\quad {\rm for}\quad (A,B) ~=~ (0,4),(j,k), &\herm a\cr
&X^{\dag}_{AB} ~=~ X_{AB}\quad {\rm for}\quad (A,B) ~=~ (0,j),(j,4), &\herm b\cr}$$
where ~$j,k ~=~ 1,2,3$~. Note that the four
generators in \herm{a} (or \fd{a})  are compact. They span the maximal compact subalgebra ~$\ck ~=~
so(3) \oplus so(2)$, the $so(3)$ being spanned by the generators ~$X_{jk}\,$, $j,k=1,2,3$,
the $so(2)$ being spanned by the generator $X_{04}\,$. The six generators in \herm{b}
(or \fd{b}) are non-compact.

Thus, in this basis, we can identify - up to sign - the Hermitian conjugation with the
Cartan involution ~$\theta$~ which in general is defined by:
\eqn\invo{\eqalign{
 \theta : X\mapsto X ~, &\quad{\rm if}\quad X\quad {\rm is ~compact}\  ,\cr
 \theta : X\mapsto -X , &\quad{\rm if}\quad X\quad {\rm is ~non compact} ~.
                \cr}}

{}From the above we have explicitly a finite-dimensional representation for
the Cartan-Weyl basis:
\eqna\cawe
$$\eqalignno{
& H_1 =  \pmatrix{\s_3 & 0 \cr 0& \s_3}
\ , \quad X^+_1 =  \pmatrix{\s_+ & 0 \cr 0& -\s_+ }
\ , \quad X^-_1 =  \pmatrix{\s_- & 0 \cr 0& -\s_- } \ ,&\cawe a\cr
& H_2 =  \pmatrix{ e_2 & 0 \cr 0 & -e_1 }
\ ,\quad
X^+_2 =  \pmatrix{ 0 & \s_- \cr 0& 0}
\ ,\quad
X^-_2 =  \pmatrix{ 0 & 0\cr \s_+ & 0}
\ ,&\cawe b\cr
& X^+_3 =  \pmatrix{ 0 & 1_2 \cr 0& 0}
\ ,\quad
X^-_3 =  \pmatrix{ 0 & 0\cr 1_2 & 0}
\ ,\quad
H_3 = H_1 +2H_2 =\pmatrix{ 1_2 & 0 \cr 0 & -1_2 }
\ ,\qquad &\cawe c\cr
&X^+_4 =  \pmatrix{ 0 & \s_+ \cr 0& 0}
\ ,\quad
X^-_4 =  \pmatrix{ 0 & 0\cr \s_- & 0}
\ ,\quad
H_4 = H_1 +H_2 =\pmatrix{ e_1 & 0 \cr 0 & -e_2 }
\ ,\qquad &\cawe d\cr
& e_1 \equiv \ha(1+\s_3) = \pmatrix{ 1 & 0\cr 0 & 0}\ ,\quad
 e_2 \equiv \ha(1-\s_3) = \pmatrix{ 0 & 0\cr 0 & 1}\ ,\cr
& \s_+ \equiv \ha (\s_1  + i \s_2) = \pmatrix{ 0 & 1\cr 0 & 0}\ ,\quad
 \s_- \equiv \ha (\s_1  - i \s_2) = \pmatrix{ 0 & 0\cr 1 & 0}\ .\quad
}$$

\newsubsec{Structure theory.}~~~
We need some more structure theory related to the applications to conformal field theory,
namely, we need the following (Bruhat) decomposition
\eqn\bruh{\cg ~=~ \cn_+ \oplus \cm \oplus \ca \oplus \cn_-}  in which all
four subalgebras have physical meaning, namely, the subalgebra ~$\cm$~
is the Lorentz subalgebra of three-dimensional Minkowski space-time $M^3$, i.e., $\cm = so(2,1)$,
the subalgebras $\cn_+,\cn_-$ are  abelian and represent the translations of $M^3$ and special conformal
transformations of $M^3$, and the algebra $\ca$  represents the dilatations of $M^3$ ($\ca$ commutes with $\cm$).
Explicitly, we have:
\eqn\basbru{
\cm ~:~ \{ X_{12}\,, ~ X_{a3}\,, ~a=1,2 \}
\,, \quad \ca ~:~ \{  H_3 \}\,,  \quad
\cn_\pm ~:~ \{ X^\pm_k \ , ~~~k = 2,3,4 \}\ . }
Note that the generators ~$H_1\,, ~ X^\pm_1$~ are complex linear combinations of those of $\cm$
(cf. \crab{a}), i.e., span the complexification ~$\cm^\bac = so(3,\bbc)$~ of $\cm$,
and actually represent a triangular decomposition of ~$\cm^\bac ~=~
\cm^\bac_+ \oplus \cm_h \oplus \cm^\bac_-$~: ~$X^\pm_1$ spanning $\cm^\bac_\pm\,$,
~$H_1$~ spanning the Cartan subalgebra $\cm_h\,$. Matters are arranged so that the factors
from \deca{} are related to the complexification of \bruh{} in the following obvious manner:
\eqn\bruhc{ \cg_\pm ~=~ \cn_\pm \oplus \cm^\bac_\pm \ , \quad \ch ~=~ \ca \oplus \cm_h \ .}

\newsubsec{Lie groups.} ~~~Finally, we introduce the corresponding connected Lie groups:
~$G ~=~ SO_0(3,2)$ with Lie algebra ~$\cg=so(3,2)$,
~$K ~=~ SO(3) \times SO(2)$~  is the maximal
compact subgroup of $G$, ~$A ~=~ \exp (\ca) ~=~
SO_0(1,1)$~ is abelian simply connected,
~$N_\pm ~=~ \exp (\cn_\pm)$~ are abelian simply
connected subgroups of ~$G$~ preserved by the action of ~$A$.
The group  ~$M ~\cong~ SO_0(2,1)$~ (with Lie algebra $\cm$)
commutes with $A$.
The subgroup ~$P = MAN$, where $N=N_+$ or $N=N_-$, is a  ~{\it maximal parabolic
subgroup}~ of $G$.
Parabolic subgroups are important because the representations
induced from them generate all admissible irreducible
representations of $G$ \La,\KZ.

\np

\newsec{Representations and invariant operators}

\newsubsec{Elementary representations.}~~~
We use the approach of \Doa{} which we adapt in a condensed form here.
We work with  so-called ~{\it elementary representations} (ERs).
They are induced from representations of ~$P ~=~ MAN_+\,$, with the factor $N_+$ being represented trivially.
We take  ~$p=0,1,2,\ldots$~ to fix a
~$(p+1)$-dimensional representation   of ~$M$,
and ~$c\in\bbc$~ to fix a (non-unitary)
character of $A$. This data is enough to determine a weight
~$\L\in\ch^*$~ as follows: ~$\L(H_1) = p$, ~$\L(H_3)  ~=~ c\,$.
Thus, we shall denote the ERs by ~$C^\L$. They can be
considered also  ~{\it holomorphic} elementary
representations of ~$G^\bac$~ and their functions can be taken to be
complex-valued ~$C^\infty$~ functions on $G$ or  $G^\bac$.
The representation action is given by the ~{\it left regular action}, which
infinitesimally is:
\eqn\ral{ (\pi_\L(X)\,\vf) (g) ~\doteq~ {d\over dt} \vf
(\exp(-t X)g)\vr_{t=0}   }
where $X\in\cg$, $g\in G$, and these can be extended
to $\cgc$ and $G^\bac$. These functions possess the
properties of right covariance \Doa{} which here means:
\eqna\low
$$\eqalignno{& {\hat X} \vf ~~=~~ \L(X) \cdot \vf ~,
\quad X \in \ch
&\low a\cr
&{\hat X} \vf ~~=~~ 0 ~, \quad X \in \cg_+   &\low b\cr }$$
where ~${\hat X}$~ is the
~{\it right}~ action of the generators of the algebra $\cg$
\eqn\rac{ ({\hat X}\vf) (g) ~\doteq~ {d\over dt} \vf
(g\exp(t X))\vr_{t=0}   }
Right covariance is used also to pass from
functions on the group $G$ to the so-called ~{\it reduced}
functions ~$\hp$~ on the coset space
~$G^\bac/B$, where ~$B ~=~ \exp( \ch) \exp (\cg_+)$~ is a Borel
subgroup of $G^\bac$, (and we use the fact that the group $G$ is maximally split
and we can work with $G^\bac$ instead).
Note that $G^\bac/B$ is a completion of
~$G_- = \exp (\cg_-)$~ and as usually we shall use  the
local coordinates of ~$G_-$~:
\eqn\axpo{ G_- ~=~ \Bigl\{ g = \pmatrix{
1 & 0 & 0 & 0 \cr
z & 1 & 0 & 0 \cr
\xi + z v/2 & v & 1 & 0 \cr
\eta - vz^2/6 & \xi - z v/2 & -z & 1 \cr} \Bigr\} }
obtained from exponentiation of the general  term: ~$z X^-_1 + v X^-_2 +\xi X^-_3 + \eta X^-_4\,$~
of ~$\cg_-\,$. The functions ~$\hp(z,v,\xi,\eta)$~ are polynomials in the variable ~$z$~ of degree ~$p$~
and smooth functions in the other three variables.
Consistently with the above, we have:
\eqna\rcoh
$$\eqalignno{ & {\hat H}_1\, \hp ~=~ p\, \hp \ , ~~p=0,1,\ldots \ , &\rcoh a\cr
& {\hat H}_3\, \hp ~=~ c\, \hp \   .&\rcoh b\cr}$$
The right action of ~$\cg_-$~ on $\hp$ is calculated easily and after some changes
of variables ~$(z,v,\xi,\eta) \mt (z,v,x,u)$~ we obtain:
\eqn\racta{ \hx^-_1  ~=~ \pd_z  \ ,
~~~\hx^-_2  ~=~ \pd_v - z \pd_x + z^2 \pd_u\ ,
~~~\hx^-_3  ~=~ \pd_x - 2 z \pd_u \ ,
~~~\hx^-_4  ~=~ \pd_u  \ .}

The left, representation, action on ~$\hp$~ is derived in a straightforward way and
after the same changes of variables we have:
\eqna\lacty
$$\eqalignno{
\pi_\L (X^+_1) ~=&~\ z^2\pd_z + 2x\pd_v + u \pd_x  - z p&\lacty a\cr
\pi_\L (X^+_2) ~=&~ -(x+ z v)\pd_z +  v^2 \pd_v +
v x  \pd_x + x^2 \pd_u - \ha (c-p) v &\lacty b\cr
\pi_\L (H_1) ~=&~\ 2z\pd_z - 2 v \pd_v + 2 u\pd_u - p &\lacty c\cr
\pi_\L (H_2) ~=&~ - z\pd_z + 2 v \pd_v +  x\pd_x - \ha (c-p) &\lacty d\cr
\pi_\L (X^-_1) ~=&~ -\pd_z +  v \pd_x + 2x \pd_u &\lacty e\cr
\pi_\L (X^-_2) ~=&~ -\pd_v    &\lacty f\cr
}$$

In addition, we have:
$$\eqalignno{
\pi_\L (X^+_3) ~=&~ [ \pi_\L (X^+_1), \pi_\L (X^+_2)] ~=~ (v z^2 -u)\pd_z + 2xv\pd_v + (x^2 +u v)\pd_x
+\cr &+ 2x u\pd_u - (x +z v) p - (c-p) x &\lacty g\cr
\pi_\L (X^-_3) ~=&~ [ \pi_\L (X^-_2), \pi_\L (X^-_1)] ~=~ - \pd_x &\lacty h\cr
\pi_\L (H_3) ~=&~ [\pi_\L (X^+_3), \pi_\L (X^-_3)] ~=~ \pi_\L (H_1)  +2 \pi_\L (H_2) ~=\cr
~=&~ 2x\pd_x + 2 v \pd_v + 2 u\pd_u  - c &\lacty i\cr
\pi_\L (X^+_4) ~=&~ \ha [ \pi_\L (X^+_1), \pi_\L (X^+_3)] ~=~ (x z^2 + u z)\pd_z + x^2\pd_v +  x u \pd_x
+\cr &+  u^2 \pd_u - (x z+ u) p -\ha (c-p) u  &\lacty j\cr
\pi_\L (X^-_4) ~=&~ [ \pi_\L (X^-_3), \pi_\L (X^-_1)] ~=~ - \pd_u &\lacty k\cr
\pi_\L (H_4) ~=&~ [\pi_\L (X^+_4), \pi_\L (X^-_4)] ~=~ \pi_\L (H_1)  + \pi_\L (H_2) ~=\cr
~=&~ z\pd_z +  x \pd_x + 2 u\pd_u  - \ha (c+p)  &\lacty l\cr
}$$

Since the left and right actions commute we can calculate the value of the
second order Casimir in either of them and obtain:
\eqn\casil{ \pi_\L (\cc_2) ~=~ \hat{\cc}_2 ~=~
\qrt \( (p+1)^2 + (c+3)^2 \) \ . }

\nt {\it Remark:} ~~~
Note that starting from a constant function $\hp = C$ and taking $c=p$ we obtain finite-dimensional
representation consisting of polynomials in $z$ obtained by the action of $\pi_\L (X^+_1)$~:
\eqn\fdr{ (\pi_\L (X^+_1))^k \cdot C~=~ \cases{ (-z)^k {p\,!\over (p-k)!}\, C & ~~~$k\leq p$ \cr
0 & ~~~$k>p$ }}
The same would be true if $p=0$ and $\ha(c-p)$ would be a non-negative integer $r$, then we would
obtain finite-dimensional
representation consisting of polynomials in $v$ obtained by the action of $\pi_\L (X^+_2)$~:
\eqn\fdra{ (\pi_\L (X^+_2))^k \cdot C~=~ \cases{ (-v)^k {r\,!\over (r-k)!}\, C & ~~~$k\leq r$ \cr
0 & ~~~$k>r$ }}

\newsubsec{Verma modules and singular vectors.}~~~
We note that conditions \low{} are the defining conditions
for the highest weight vector of a highest weight module (HWM)
over $\cgc$ with highest weight $\L$, in particular, of a Verma
module with this highest weight.
Let us recall that a ~{\it Verma module} ~$V^\L$~ is defined as
the HWM over ~$\cgc$~ with highest weight ~$\L \in \ch^*$~ and
highest weight vector ~$v_0 \in V^\L$, induced from the
one-dimensional representation ~$V_0 \cong \bbc v_0$~ of
~$U(\cb)$~, where ~$\cb  = \ch \oplus \cg_+$~ is a Borel
subalgebra of ~$\cg$, such that:
\eqn\indb{
\eqalign{ &X ~v_0 ~~=~~ 0 , \quad \forall\, X\in \cg_+ \cr
&H ~v_0 ~~=~~ \L(H)~v_0\,, \quad \forall\, H \in \ch \cr }}

Verma modules are generically irreducible. A Verma  module ~$V^\L$~ is
reducible \BGG{} iff there exists a root ~$\b \in\D^+$~ and ~$m\in\bbn$~
such that
\eqn\red{ (\L + \r~, ~\b^\vee) ~=~  m }
holds, where ~$\r = {1 \over 2}\sum_{\a \in \D^+}~\a$~.
If \red{} holds then $V^\L$ is reducible. It contains an invariant submodule
which is also a Verma module ~$V^{\L'}$~ with shifted weight ~$\L'=\L-m\b$.
This statement is equivalent to the fact that $V^\L$ contains a
~{\it singular vector}~
~$v_s \in V^\L$, such that ~$v_s ~\neq ~\xi v_0\,$, ($0\neq\xi\in\bbc$),
and~:
\eqn\sing{
\eqalign{& X ~v_s ~~=~~ 0 , \quad \forall\, X\in \cg_+ \cr
&H ~v_s ~~=~~ \L'(H) ~v_s\,, \quad
\L' ~=~ \L - m\b, ~~\forall\,
H \in \ch \cr }}

It is important that one can find explicit formulae for the singular
vectors. The singular vectors introduced above are given by \Doa~:
\eqn\siv{ v_s ~=~ v^{\a,m_\a} ~=~ {\cal P}^{\a,m_\a} (\cg_-) ~v_0 }
where ~${\cal P}^{\a,m_\a}$~ is a homogeneous polynomial in the
generators of $\cg_-$ of weight ~$\a\,m_\a\,$~  and is unique up to
a non-zero multiplicative constant.   The
conditions \red{} spelled out for the four positive roots in our
situation are: \eqna\reda
$$\eqalignno{
&m_1 ~=~ m_1(\L) ~\doteq~  \L(H_1) + 1  ~\in\bbn \ , &\reda a\cr
&m_2 ~=~ m_2(\L) ~\doteq~ \L(H_2) + 1   ~\in\bbn \ ,   &\reda b\cr
&m_3 ~=~ m_3(\L) ~\doteq~ \L(H_3) + 3 = m_1 + 2m_2   ~\in\bbn \ , &\reda c\cr
&m_4 ~=~ m_4(\L) ~\doteq~  \L(H_4) + 2 = m_1 + m_2  ~\in\bbn \ .
&\reda d\cr}$$
The singular vectors corresponding to these cases   are~:
\eqna\svsv
$$\eqalignno{  &v^{\a_1,m_1} ~=~  (X^-_1)^{m_1}\, v_0 ~,  \quad m_1\in\bbn\ ,
&\svsv a\cr
&v^{\a_2,m_2} ~=~  (X^-_2)^{m_2}\, v_0 ~, \quad m_2\in\bbn\ , &\svsv b\cr
&v^{\a_3,m_3} ~=~  \sum_{k=0}^{m_3}\ a_k\
(X_2^-)^{m_3-k}\, (X_1^-)^{m_3} \, (X_2^-)^{k}\, v_0 ~,
\quad m_3\in\bbn\ ,
&\svsv c\cr
&a_k ~=~ \cases{ a_0\, (-1)^k\, {m_3 \choose k} \, { m_2 \over
m_2 -k}\ , ~~~&~~ $m_2 \notin \{\, 0,\ldots, m_3\,\}$  \cr
a_0\ \d_{k,m_2} \ , ~~~&~~ $m_2 \in \{\, 0,\ldots, m_3\,\}$
} &\svsv {c'}\cr
&v^{\a_4,m_4} ~=~  \sum_{k=0}^{2m_4}\ b_k\
(X_1^-)^{2m_4-k}\, (X_2^-)^{m_4} \, (X_1^-)^{k}\, v_0 ~,
\quad m_4\in\bbn\ , &\svsv d\cr
&b_k ~=~ \cases{ b_0\, (-1)^k\, {2m_4 \choose k} \, { m_1  \over
m_1 -k}\ , ~~~&~~ $  m_1 \notin \{\, 0,\ldots, 2m_4\,\}$  \cr
b_0\ \d_{k,m_1} \ , ~~~&~~ $m_1 \in \{\, 0,\ldots, 2m_4\,\}$}
\qquad &\svsv {d'}
}$$
(Note that in each of the four cases \svsv{} only the relevant $m_j$ must
be a natural number (as displayed).)
Formulae \svsv{a,b} are the general expressions valid for any simple root \Doa,
while \svsv{c,d} are given in \Dosv.
Certainly, \red{} may be fulfilled for several positive roots, even
for all of them if \red{} is fulfilled for all simple roots.

For future use we note one important special case of \svsv{c} when ~$m_3$~ is even,
and ~$m_1=1$, then ~$m_2 = \ha (m_3-1)\notin\bbz$. In this case
\svsv{c} can be written
as follows:
\eqn\svsvc{ v^{\a_3,m_3} ~=~ \( (X^-_3)^2 - 4 X^-_4 X^-_2 \)^{\ha m_3}\ v_0 ~+~ P(\cg_-)\ X^-_1\ v_0 \ ,
\qquad m_3\in2\bbn, ~m_1=1 \ , }
where in order to achieve this simple form we have introduced also the non-simple root vectors
~$X^-_3,  X^-_4\,$, and ~$P(\cg_-)$~ is a polynomial whose exact form is not important
for our purposes.

Note that independently from the question of reducibility the numbers ~$m_1,m_2$~
may be used to characterize the representations $C^\L$ and $V^\L$.
In particular, the second order Casimir is expressed in a simple manner through
these numbers after evaluation on the  highest weight vector:
\eqna\casis
$$\eqalignno{
\cc_2\, v_0 ~=&~ \{\,
\ha (X^+_1  X^-_1 + X^-_1  X^+_1) \ +\
X^+_2  X^-_2 + X^-_2  X^+_2 \ +\
\ha (X^+_3  X^-_3 + X^-_3  X^+_3)
\ +\cr &+~ X^+_4  X^-_4 + X^-_4 X^+_4 \ +\
\ha H_1^2 \ +\  H_2^2 \ +\  H_1 H_2\, \}\, v_0 ~=\cr
=&~ \{\,
\ha H_1  \ +\ H_2  \ +\ \ha H_3   \ + H_4   \ +\
\ha H_1^2 \ +\  H_2^2 \ +\  H_1 H_2\, \}\, v_0 ~=\cr
=&~ \(\, \ha m_1^2 + m_2^2 + m_1m_2\, \)\, v_0 ~=&\casis a\cr
=&~ \qrt \(\,  m_1^2 + m_3^2 \, \)\, v_0 ~=&\casis b\cr
=&~ \ha \(\,  m_2^2 + m_4^2 \, \)\, v_0\ .&\casis c\cr}$$
(Passing from the first to the second equality we use the commutation relations
and the defining properties of $\ v_0\,$,
e.g., ~$(X^+_1  X^-_1 + X^-_1  X^+_1)\, v_0 ~=~ X^+_1  X^-_1  \, v_0 ~=~
H_1  \, v_0\,$.) Comparison with \casil{} is straightforward if we substitute in \casis{b}
~$m_1 = \L(H_1)+1 = p+1$, $m_3  = \L(H_3)+3= c+3$.

It is useful for future reference to write down the numbers corresponding to
the four cases of ~$\L'$, i.e., to write down ~$m_1(\L'),m_2(\L')$~ for the four cases
of \reda{}. In fact, though it is redundant, the picture becomes more instructive,
if we write down also ~$m_3(\L'),m_4(\L')$.
Explicitly, we have for $m_i \equiv m_i(\L)$~:
\eqna\chan
$$\eqalignno{ \( m_1(\L'),m_2(\L'),m_3(\L'),m_4(\L') \) ~=&~
\( -m_1,m_4,m_3,m_2\) \ , \cr
&~ \L' = \L -m_1\a_1, ~~m_1\in\bbn \ , & \chan a\cr
\( m_1(\L'),m_2(\L'),m_3(\L'),m_4(\L') \) ~=&~
\( m_3,-m_2,m_1,m_4\) \ , \cr
&~ \L' = \L -m_2\a_2, ~~m_2\in\bbn \ , & \chan b\cr
\( m_1(\L'),m_2(\L'),m_3(\L'),m_4(\L') \) ~=&~
\( m_1,-m_4,-m_3,-m_2\) \ , \cr
&~ \L' = \L -m_3\a_3, ~~m_3\in\bbn \ , & \chan c\cr
\( m_1(\L'),m_2(\L'),m_3(\L'),m_4(\L') \) ~=&~
\( -m_3,m_2,-m_1,-m_4\) \ , \cr
&~ \L' = \L -m_4\a_4, ~~m_4\in\bbn \ . & \chan d\cr
}$$
Since the Verma modules ~$V^{\L'}$~ are isomorphic to submodules of ~$V^{\L}$~
they have the same Casimirs. For ~$\cc_2$~ this is most obvious from \casis{b,c}.

All the above is valid for Verma modules also over ~$so(5,\bbc)$.
Now we shall express conditions \red{} for the four positive roots
taking into account the signatures of our representations   (cf. \DoSe):
\eqna\redas
$$\eqalignno{ &m_1 =  \L(H_1) + 1 = 2 s_0 + 1~ , \quad s_0 =\ha p\, \in \ha\bbz_+ \ ,
&\redas a\cr &m_2 = \L(H_2) + 1 = 1 - E_0 - s_0 ~ , \quad E_0 = -\ha c \ ,
&\redas b\cr &m_3 = \L(H_3) + 3 = m_1 + 2m_2 ~=~ 3 - 2E_0 ~ ,
&\redas c\cr
&m_4 =  \L(H_4) + 2 = m_1 + m_2 = 2 - E_0 + s_0 ~
, &\redas d\cr}$$
where we have introduced also the traditionally used energy $E_0$ and  spin $s_0\,$.
For future reference we record an explicit expression for ~$\L$~:
\eqn\expli{ \L ~=~ \ha (m_1-1)\, \a_1 ~+~ \ha (m_3-3)\, \a_3 ~=~
s_0\a_1 ~-~ E_0\a_3 ~. }

The eigenvalue of ~$H_3$~ is called the energy since upon contraction of
~$so(3,2)$~ to the Poincar\'e algebra, the generator ~$H_3$~ goes
to the translation operator ~$P_0\,$.
Analogously, ~$H_1$~ is the third component of the angular momentum.
In terms of ~$E_0,s_0$~  the Casimir has the expression:
\eqn\casise{
\cc_2 ~=~ E_0 (E_0-3) + s_0(s_0+1) + 5/2 }
which (up to the additive constant) was the preferred usage in \FFa,\FFb,\FFc.
(In the latter papers the irreducible representations of ~$so(3,2)$~ are
denoted by ~$D(E_0,s_0)$.)

Next we note that if ~$m_1, m_2\in{\bbn}$~, then the irreducible
representations with non-positive energy  ~$E_0 ~=~ (3-m_1-2m_2)/2$, and spin ~$s_0 ~=~
(m_1-1)/2$, are the non-unitary finite-dimensional representations of ~$\cg$. Their
dimension is: ~$m_1m_2m_3m_4/6$. (For example
~$m_1 ~ m_2 =1$~ gives the trivial 1--dimensional
representation; the fundamental representations are obtained for ~$m_1 = 1,
~m_2 = 2$~ and ~$m_1 = 2, ~m_2 = 1\,$.)

But we are interested in the positive energy UIRs
 of ~$\cg$~ given as follows \refs{\Dirac--
 \Evans} \
(with $s_0\in\ha\bbz_+$):
\eqna\unita
$$\eqalignno{
& {\rm Rac} ~:~ D(E_0,s_0) ~=~ D(1/2,0) ~,
\qquad {\rm Di} ~:~ D(E_0,s_0) ~=~ D(1,1/2) ~, &\unita a\cr
&D(E_0 > 1/2 , ~s_0\, =\, 0) ~,
\quad D(E_0 >1 , ~s_0 \, =\, 1/2) ~,
\quad D(E_0 \geq s_0 + 1 , ~s_0\,\geq\, 1) .\qquad  &\unita b}$$
The first two are the singleton representations discovered by Dirac \Dirac{}
and the last ones for ~$E_0 = s_0+1$~  correspond to the spin-$s_0$~ massless
representations. ~~~
Comparing the list \redas{} with \unita{} we note that \redas{a} holds
in all cases of \unita{}, while \redas{b} never holds because ~$m_2\leq
1/2$~. Next, we note that ~$m_3$~ is a positive integer only for ~$E_0 ~=~
1/2, 1$, in which case ~$m_3 ~=~ 2,1$, respectively. Similarly, ~$m_4$~ is a
positive integer only for ~$E_0 -s_0 ~=~ 1$, and that integer is ~$m_4 ~=~
1$.

The singular vectors corresponding to these cases are (choosing the normalization
constants appropriately):
\eqna\sv
$$\eqalignno{
v^{\a_1,m_1} ~=&~ (X^-_1)^{m_1} ~v_0  ~, ~~m_1 = 2s_0+1\in\bbn ~,
&\sv a\cr
v^{\a_3,1} ~=&~ (s_0\, X^-_3 +  X^-_2 X^-_1) ~v_0 ~,
~~m_3 =1 ~, &\sv b\cr
v^{\a_3,2} ~=&~
\{ (2s_0-1)(2s_0+3)\, X_2^-\, (X_1^-)^2\,X_2^-
~-~ \ha (2s_0+1)(2s_0+3)\,(X_2^-)^2\, (X_1^-)^2
~-\cr
&-~ \ha (2s_0-1)(2s_0+1)\,(X_1^-)^2\, (X_2^-)^2 \} v_0 ~=\cr
=&~ \{\, (1-4s^2_0)\,(X^-_3)^2 ~-~ 4 (1-2s_0)\, X^-_4\,X^-_2 ~+\cr
&+~ 4 (1-2s_0)\, X^-_3\, X^-_2\, X^-_1 ~-~ 4(X_2^-)^2\, (X_1^-)^2
\ \}\ v_0 ~,
~~~~m_3=2\ ,   &\sv c\cr
v^{a_4,1} ~=&~ \{\, 2s_0 (2s_0-1)\, X^-_4 +  (2s_0-1)\,
X^-_3\, X^-_1 +  X^-_2\, (X^-_1)^2\ \}\ v_0 ~, ~~~m_4=1 ~, &\sv d\cr}$$
where we have introduced also the non-simple root vectors (choosing the
appropriate ordering).
Note that \sv{b} for $s_0 =0$, ~\sv{c} for $s_0 =1/2$,
and \sv{d} for $s_0 =0,1/2$ are composite singular vectors being
descendants of \sv{a}. (The latter follow also from the general formulae
\svsv{}: for \sv{b,c} from \svsv{c} with $m_2=\ha(m_3-m_1)=0$,
and for \sv{d} from \svsv{d} with $m_1 =1,2 \in \{0,1,2=2m_4\,\}$.)

\newsubsec{Invariant differential operators.}~~~
The main ingredient of the procedure of \Doa{} is that to every
singular vector  there corresponds an invariant differential operator.
Namely, to the singular vector
~$v_s ~=~ v^{\a,m_\a}$~ (cf. \siv) of the Verma module ~$V^\L$~
there corresponds an invariant differential operator
\eqn\intt{D^{\a,m_\a} ~:~ C^\L ~\lra ~C^{\L -m_\a \a} }
given explicitly by:
\eqn\dvv{ D^{\a,m_\a} ~=~ {\cal P}^{\a,m}
(\hat{\cg}_-) } where
~${\cal P}^{\a,m}$~ is the same polynomial as in \siv,
and ~$\hat{\cg}_-$~ symbolizes the right action \rac.

These operators give rise to the $\cg$-invariant equations:
\eqna\ine
$$\eqalignno{ &D^{\a,m_\a}\, \hp ~=~ \hp'\ , \quad
\hp\in C^\L, ~~\hp'\in C^{\L -m_\a \a} , \quad \a\ {\rm non-compact} &\ine a\cr
&D^{\a,m_\a}\,\hp~=~ 0 ~, \quad
\hp\in C^\L,  \quad \a\ {\rm compact} &\ine b\cr}$$
We recall that a compact root is defined by the property that it has zero value
on the dilatation subalgebra ~$\ca$, i.e., in our case, on the generator ~$H_3\,$,
which means that only the root $\a_1$ is compact.
In fact, equation \ine{b} just expresses the fact that we have induction from
finite-dimensional irreps of the Lorentz group $M$. In our case, the counterpart
of \ine{b} is: ~$\pd_z^{m}\,\hp ~=~ 0$, which is trivially satisfied since all our functions
are polynomials of degree $2s_0 = m-1= p$.

The representations in \ine{a} have the same Casimir. The operators ~$D^{\a,m_\a}$~
have the intertwining property:
\eqn\inte{ D^{\a,m_\a}\,\circ\, \pi_\L(X)\, \hp ~=~ \pi_\tL(X)\,\circ\, D^{\a,m_\a}\, \hp \ ,
\qquad \tL=\L -m_\a \a, ~~
\forall X\in\cg, ~~\forall \hp\in C^\L  }
In spite of \inte{} the ERs  ~$C^\L$~ and ~$C^\tL$~ are not equivalent
but only ~{\it partially}~ equivalent since the differential operators $D^{\a,m_\a}$
have   nontrivial kernels and/or images (for more detailed explanations of these notions
we refer to \DMPPT).

For future use we mention one important case in which arises the
d'Alembert operator. Namely, we consider ~$m_1=1$, $m_3\in 2\bbn$, the singular vectors
are \svsv{a} and \svsvc{} (the latter equivalent to \svsv{c}). Thus, the relevant equations  are:
\eqna\raccc
$$\eqalignno{  \pd_z\, \hp ~=&~ 0\ ,  \qquad m_1=1\ ,&\raccc a\cr
D^{\a_3,m_3}\ \hp ~=&~
\{ ( \pd_x - 2 z \pd_u)^2 ~-~ 4 \,\pd_u\, (\pd_v - z \pd_x + z^2 \pd_u) \}^{\ha m_3}\, \hp
~+~ P(\hat{\cg}_-)\ \pd_z\, \hp ~=\cr
=&~ \( \pd_x^2\ -\ 4 \pd_u \pd_v\)^{\ha m_3}\ \hp   ~=~ \hp'\ ,
\qquad m_1=1, ~m_3\in 2\bbn\,.\qquad
 &\raccc b\cr}$$

Note that  the operator ~$\pd_x^2\ -\ 4 \pd_v    \pd_u$~ is the d'Alembert operator
in $M^3$ if we identify the coordinates as follows:
\eqn\mink{ x ~=~ y_0\ , \quad u ~=~ y_1 - iy_2
\ , \quad v ~=~ y_1 + iy_2 }
Then indeed we have:
\eqn\dal{\pd_y^2\ -\ 4 \pd_v    \pd_u ~=~ \pd^2_0 - \pd^2_1 - \pd^2_2
~\equiv~ \Box\ .}
Analogously, the Minkowski length is given by:
\eqn\minkl{ x^2 - uv ~=~ y_0^2 - y_1^2 - y_2^2 ~\equiv ~ y^2 \ .}

\newsubsec{Quartet of reducible representations.} ~~~Here we treat
a quartet of reducible representations which is very important since it contains all
finite-dimensional irreps of $so(3,2)$. To be more explicit we start with $C^\L$ such
that ~$m_j \equiv m_j(\L)$~ are natural numbers. Then we have the following diagram:
\eqna\diagr
$$\eqalignno{
&\matrix{ C^\L & \llra{3,m_3} & C^{\L-m_3\a_3} \cr
&&\cr
\downa && \upa\cr
&&\cr
C^{\L-m_2\a_2} & \llra{3,m_1} & C^{\L-m_2\a_2-m_1\a_3} } & \diagr{a}}$$
where the numbers at the arrows denote the corresponding differential operators
$D^{\a_k,m'_k}$, ~$k=2,3,4$.
It is also very instructive to write the same diagram with explicit signatures:
$$\eqalignno{
& \matrix{ \(m_1,m_2,m_3,m_4\) & \llra{3,m_3} & \( m_1,-m_4,-m_3,-m_2\) \cr
&&\cr
\downa && \upa\cr
&&\cr
\( m_3,-m_2,m_1,m_4\) & \llra{3,m_1} & \( m_3,-m_4,-m_1,m_2\)} & \diagr{b}}$$

The representations in \diagr{} have the same Casimir and are partially equivalent.
It is natural to ask whether \diagr{} is a commutative diagram. Formally, it is,
since as differential operators we have:
\eqn\commd{ D^{\a_3,m_3} ~=~ D^{\a_4,m_2} \circ D^{\a_3,m_1}
\circ D^{\a_2,m_2} }
However, the kernels and images of these operators may not coincide,\foot{This question
will be discussed in more details in \DoMop.} since the operator
$D^{\a_3,m_3}$  factorizes also according to
the second line of \svsv{c'}:
\eqn\commc{D^{\a_3,m_3} ~=~ D^{\a_2,m_4} \circ D^{\a_1,m_3}
\circ D^{\a_2,m_2} }
(This second decomposition of ~$D^{\a_3,m_3}$~ does not appear on \diagr{} since it involves
   one intermediate space which is unphysical - the $z$-dependence is not polynomial.)

The finite-dimensional irreps of $so(3,2)$ shall be denoted by ~$E_{m_1,m_2}\,$.
They correspond to the holomorphic (or anti-holomorphic) irreps of $so(5,\bbc)$
and have the same dimension: ~$\dim_{m_1,m_2} ~=~ m_1m_2m_3m_4/6$. In our setting all
of these are subspaces of $C^\L$ in the top-left corner of diagram \diagr{} and are
obtained in the following way:
\eqn\fdrb{\eqalign{ E_{m_1,m_2} ~=~ \{ \ &f\in C^\L ~:~ m_k = m_k(\L) ~\in\bbn \ ,
\quad D^{\a_1,m_1}\, f ~=~ \pd_z^{m_1}\, f ~=~ 0, \cr
&D^{\a_2,m_2}\ f ~=~ 0 \ \}\ }}
Of course, the first equation is fulfilled for $C^\L$ by definition,
but we include it for completeness. The finite-dimensional subspaces are also
in the kernel of the operator ~$D^{\a_3,m_3}$~ due to decomposition \commc.

The diagram \diagr{} is important also because it contains effectively
an additional integral operator
which we shall not consider in detail but just mention at this point.
This operator is related to the so-called restricted root system of $\cg$ w.r.t. $\ca$,
where the root spaces are ~$\cn_\pm\,$.
In our case, cf. \basbru{}, this restricted root system  has just one root
which coincides with ~$\a_3$~ restricted to ~$\ca$~ due to the facts:
~$[H_3,X^\pm_1] ~=~ 0$,
~$[H_3,X^\pm_k] ~=~ \pm 2\, X^\pm_k\,$, ~$k=2,3,4$. The action of this root on the signatures
naturally coincides with the action of ~$\a_3\,$,
but the action itself is valid for arbitrary signatures. The operator realizing such an action
is an integral operator involving the conformal two-point function ~$G_\L\,$,
~$\L \cong (m_1,m_2,m_3,m_4)$. For scalar functions ($m_1=1$) the latter is   given  by:
\eqn\twop{ G_\L (y) ~=~ {N(\L) \over (y^2)^{\ha(3+m_3)}} \ , \qquad y^2 = y_0^2 - y_1^2 - y_2^2 \ ,}
where ~$N(\L)$~ is a normalization constant related to the Plancherel measure,\foot{The $y$-dependence
of $G_\L (y)$ was given first in \FGGP{} for arbitrary space-time dimension
$D$ (replace $3+m$ with $D+m$), the group theory interpretation belongs to
mathematical \KS{} and physical \Koller{} work, for the interpretation of $N(\L)$
and more details we refer to \DMPPT.} and the integral operator itself is given by:
\eqn\stein{\eqalign{ &A_\L ~:~ C^\L ~\longrightarrow~ C^\tL \ ,
\qquad \tL = \L - m_3\a_3 \ , \cr
&(A_\L\,\hp) (y) ~=~ \int \ G_\L (y-y')\ \hp(y')\  d^3y' \ , \qquad
\hp \in C^\L \ .  }}
If ~$m_3\notin\bbz$~ the operators ~$A_\L$~ and ~$A_\tL$~ are inverse to each other:
\eqn\inve{A_\tL \circ A_\L ~=~ {\rm id}_{C^\L} ~ ,
\qquad A_\L \circ A_\tL ~=~ {\rm id}_{C^\tL} ~ .}
For ~$m_3\in 2\bbn$~ the operator ~$A_\L$~ reduces to ~$D^{\a_3,m_3}$~ in the
same situation (cf. \raccc{}, \mink, \dal) due to
the following:
\eqn\gene{\eqalign{
\lim_{\eps\, \mt +0}\ G_{\L(m_3 + \eps)}(y) ~=&~
\lim_{\eps\, \mt +0}\ {N(\L(m_3 + \eps)) \over (y^2)^{\ha(3+m_3+\eps)}} ~\sim~
\lim_{\eps\, \mt +0}\ {\eps \over (y^2)^{\ha(3+m_3+\eps)} }
~\sim~ \cr &\sim~ \Box^{\ha m_3}\, \d^3(y) }}
(cf. \GeSh,\DMPPT,\DoPe) and then we have:
\eqn\redu{(A_\L\,\hp) (y) ~\sim~ \int \ \Box^{\ha m_3}\, \d^3(y-y')\ \hp(y')\  d^3y'
~=~ \Box^{\ha m_3}\, \hp(y) ~=~ D^{\a_3,m_3}\, \hp(y)\ .}

For the same values ~$m_3\in 2\bbn$~ the operator ~$A_\tL$~
remains integral and acts in the ~{\it opposite\ } direction w.r.t. ~$D^{\a_3,m_3}$~
on diagram \intt{} (in the case ~$\a=\a_3$) and diagram \diagr{} (two occurrences). This is the reason
we mention the integral operators.
(Note that \gene{} remains valid also for ~$m_3=0$~ when ~$\L'=\L$~ and the
operator ~$A_\L$~ reduces to the identity operator ~${\rm id}_{C^\L}\,$.)

\newsec{Invariant differential operators and equations related to UIRs}

\nt
The differential operators related to
the unitary cases correspond to the singular vectors in  \sv{} and are given as follows:
\eqna\difo
$$\eqalignno{
D^{\a_1,m} ~=&~ ~\pd_z^{m}\ , \qquad m=m_1 = p+1 = 2s_0+1\in\bbn ~,
&\difo a\cr
D^{\a_3,1} ~=&~ ~s_0\, (\pd_x - 2 z \pd_u)\  +\ (\pd_v - z \pd_x + z^2 \pd_u)\, \pd_z\ ,
\qquad m_3 =1 ~, &\difo b\cr
D^{\a_3,2} ~=&~
(1-4s^2_0)\,(\pd_x - 2 z \pd_u)^2 ~-~ 4 (1-2s_0)\,
\pd_u\, (\pd_v - z \pd_x + z^2 \pd_u)
 ~+ \cr
&+~ 4 (1-2s_0)\,(\pd_x - 2 z \pd_u)\,(\pd_v - z \pd_x + z^2 \pd_u)\, \pd_z
 ~-\cr &-~ 4
(\pd_v - z \pd_x + z^2 \pd_u)^2\, \pd_z^2 \ , \qquad m_3=2 ~, &\difo c\cr
D^{a_4,1} ~=&~ 2s_0 (2s_0-1)\, \pd_u\ +\  (2s_0-1)\, (\pd_x - 2 z \pd_u)\,\pd_z
\ +\cr &~+\ (\pd_v - z \pd_x + z^2 \pd_u)\, \pd_z^2 \ , \qquad m_4 =1 \ .
 &\difo d\cr}$$

Next we  analyze the  important cases separately.

\newsubsec{Rac.} ~~In this case the energy and spin are: $E_0=\ha$, $s_0=0$.
The equations are \difo{a,c} (cf. also \raccc):
\eqna\racc
$$\eqalignno{  &\pd_z\, \hp ~=~ 0\ , &\racc a\cr
&D^{\a_3,2}\ \hp ~=~
\{\ (\pd_x - 2 z \pd_u)^2 ~-~ 4 \,
\pd_u\, (\pd_v - z \pd_x + z^2 \pd_u)
 ~+ &\racc {b'}\cr
&+~ 4 \,(\pd_x - 2 z \pd_u)\,(\pd_v - z \pd_x + z^2 \pd_u)\, \pd_z
~-~ 4
(\pd_v - z \pd_x + z^2 \pd_u)^2\, \pd_z^2\, \}\, \hp ~=\qquad &\racc {b''}\cr
&=~ \( \pd_x^2\ -\ 4 \pd_u \pd_v\)\ \hp ~=~ \Box\, \hp ~=~ \hp'\ .
 &\racc b\cr}$$
Note that the target space $C^{\L'}$, $\L'=\L-2\a_3\,$, is again of
scalar functions ($s'_0=0$) and unitary ($E'_0=\fha$). Thus,
$C^{\L'}$ is reducible, since the image ~$\( \pd_x^2\ -\ 4 \pd_u
\pd_v\)\ C^\L$~ can not be onto $C^{\L'}$, the functions $\hp'$ on
the RHS belong to a proper subspace of $C^{\L'}$. Note that the
Verma module $V^{\L'}$ is reducible only w.r.t. \redas{a} so that
~$m'_1=m_1=1$, thus, the functions of $C^{\L'}$   obey only one
differential equation: $~\pd_z\, \hp' ~=~ 0$.

The Rac case is the lowest case of the subfamily in which the invariant operator
is directly related to the d'Alembertian, cf. \raccc.

\newsubsec{Di.} ~~In this case the energy and spin are: $E_0=1$, $s_0=\ha$. The equations are \difo{a,b}:
\eqna\dicc
$$\eqalignno{  &\pd^2_z\, \hp ~=~ 0\ , &\dicc a\cr
&D^{\a_3,1}\ \hp ~=~ \{\ \ha\, (\pd_x - 2 z \pd_u)\  +\ (\pd_v - z
\pd_x + z^2 \pd_u)\, \pd_z\  \}\, \hp ~=~ \hp' \ . &\dicc b\cr}$$
 The invariant equation
\dicc{b} has the advantage over the usual writing of the equation
for the "Di" field (cf. \Frob) since it is a scalar equation
encompassing all information. It is also more easy to decompose the
equation in components by setting \eqn\dico{ \hp_{\rm Di} ~=~ \hp_0
\ +\ z\,\hp_1 \ , \qquad \pd_z \,\hp_i =0 \ ,} and analogously for
$\hp'$ in \dicc{b}. Substituting the above in \dicc{b} and
extracting the coefficients of the resulting polynomial in $z$ we
have: \eqna\dicco
$$\eqalignno{
&  \ha\, \pd_x \,\hp_0 ~+~ \pd_v\,\hp_1  ~=~ \hp'_0\ , &\dicco a\cr
&
  \pd_u\,\hp_0 ~+~ \ha\, \pd_x\, \hp_1 ~=~ \hp'_1 \ . &\dicco b\cr}$$
The first equation is the coefficient at $z^0$ of \dicc{b},
the second is at $z^1$, while the coefficient at $z^2$ is identically zero.

Note that the target space $C^{\L'}$~ in \dicc{b}, $\L'=\L-\a_3\,$,
is again of two-component functions ($s'_0=\ha$) and unitary
($E'_0=2$). Thus, $C^{\L'}$ is reducible, as in the Rac case, and
again the Verma module $V^{\L'}$ is reducible only w.r.t. \redas{a}
so that ~$m'_1=m_1=2$, thus, the functions of $C^{\L'}$ obey only
one differential equation: $~\pd^2_z\, \hp' ~=~ 0$.

We now consider the kernel of the differential operator ~$D^{\a_3,1}$, i.e.,
when the RHS of \dicc{} and \dicco{} are zero.
Acting on \dicco{a} by $\pd_x$ and on \dicco{b} by $-2\pd_v$ and adding together,
we obtain that $\hp_0$ fulfils the d'Alembert equation: $(\pd_x^2 - 4 \pd_v    \pd_u) \hp_0 = 0$.
We obtain the same for $\hp_1$  acting on  \dicco{b} by $\pd_x$ and on \dicco{a} by $-2\pd_u$ and adding together.
Thus, we recover the fact that the 'Di' fulfils the d'Alembert equation:
\eqn\dala{ \( \pd_x^2\ -\ 4 \pd_v    \pd_u\)\ \hp_{\rm Di} ~=~ 0\ . }
Of course, the full equations \dicc{} (or in components \dicco{}) contain more information
than \dala.

\newsubsec{Massless representations.}
~~In this case the energy and spin are: $E_0=s_0+1$,
$s_0=1,\trha,\ldots$ ~The equations are \difo{a,d}: \eqna\mass
$$\eqalignno{  &\pd^{p+1}_z\, \hp ~=~ 0\ ,  \qquad  p = 2s_0 = 2,3,\ldots \ ,
&\mass a\cr
& D^{\a_4,1}\ \hp ~=\cr &=~ \{\, p (p-1)\, \pd_u\ +\  (p-1) (\pd_x - 2 z \pd_u)\,\pd_z
\ +\ (\pd_v - z \pd_x + z^2 \pd_u)\, \pd_z^2\, \}\,\hp ~=~ \hp' \qquad
&\mass b\cr}$$
Unlike the singleton cases the target space $C^{\L'}$ is unitary but not massless:
it has $E'_0=s'_0+3\,$, ~$s'_0 = s_0 -1$, ($p'=p-2$).

Thus, from the point of view of applications to physics these invariant equations
relate a field of spin ~$s_0$~ to a field of spin ~$s_0-1$. In the simplest case
a vector field $s_0=1$ is coupled to a scalar field.

It is useful to write out \mass{b} in components using: ~$\hp ~=~ \sum_{j=0}^p\, z^j\,\hp_j\,$,
~$\hp' ~=~ \sum_{j=0}^{p-2}\, z^j\,\hp'_j\,$.
The result is ~$p-1$~ equations:
\eqn\massc{ \eqalign{ (p-j)(p-j-1)\,\pd_u\,\hp_j ~+&~
(j+1)(p-j-1)\,\pd_x\,\hp_{j+1} ~+~ (j+1)(j+2)\,\pd_v\,\hp_{j+2} ~=~ \hp'_j\ , \cr
& j=0,1,\ldots,p-2 \ .}}

Further we restrict to the kernel of ~$D^{\a_4,1}$.
In the simplest case ~$p=2$~ there is only one equation:
\eqn\massca{ 2 \pd_u\,\hp_0 ~+~ \pd_x\,\hp_1 ~+~ 2\pd_v\,\hp_2 ~=~ 0 \ ,}
which can be rewritten as equation for conserved current with the substitution:
\eqn\curr{ \hp_0 ~=~ J_1 -i J_2\ , \quad \hp_1 ~=~ - J_0 \ , \quad \hp_2 ~=~ J_1 +i J_2\ ,}
using also \mink{}:
\eqn\cons{ \pd_0\, J_0\ -\ \pd_1\, J_1\ -\ \pd_2\, J_2 ~=~ 0 \ .}

Similarly, for arbitrary ~$p=2,3,\ldots,$~ there are ~$p-1$~ independent conserved
currents ~$J^{p,j}$, ~$j=0,\ldots,p-2$,~
with components as follows:
\eqn\curra{\eqalign{
J^{p,j}_0 ~=&~ - (j+1)(p-j-1)\, \hp_{j+1} \ ,\cr
J^{p,j}_1 ~=&~ \ha \{\, (j+1)(j+2)\, \hp_{j+2} ~+~  (p-j)(p-j-1)\, \hp_{j} \,\}\ ,\cr
J^{p,j}_2 ~=&~ \iha \{\, (j+1)(j+2)\, \hp_{j+2} ~-~  (p-j)(p-j-1)\, \hp_{j} \,\}\ .\cr
}}

Finally, we would like to mention that all massless representations  appear on
diagram \diagr{} in the case ~$m_2=1$. We would like to parametrize these diagrams
by the spin of the massless irrep, i.e., by ~$s_0=1,\trha,\ldots$.
In that case the signature of the ER  $C^\L$ on the top-left corner is:
~$(2s_0-1,1,2s_0+1,2s_0)$~ and this ER contains a finite-dimensional
irrep of dimension: ~$s_0(4s^2_0-1)/3$.
The massless UIR is contained in the ER in the bottom-right corner of   \diagr{},
 while the massive UIR of spin $s_0-1$ is in the ER in the top-right corner.

\np

\newsec{Classification of ~$so(5,\bbc)$~ irreps}

\nt
In view of possible applications relating character formulae to integrability we need first
to classify the irreducible HWM over ~$so(5,\bbc)$. For this we use two important properties of
Verma modules \Dix:
\item{\bu} Every Verma module $V^\L$ contains a unique proper maximal submodule $I^\L$.
\item{\bu}  Every HWM is isomorphic to a factor-module of the Verma module with the
same highest weight. (Universality property).

Thus, among the HWM with highest weight $\L$ there is a unique
irreducible one, denoted by $L_\L $, i.e.
\eqn\irrep{ L_\L = V^\L/I^\L ~. }

Clearly, if the Verma module $V^L$ is ~{\it irreducible}~ then
~$L_\L = V^\L$. Thus, we need to classify the reducible Verma modules.

It is convenient for such classification   to use the
notion of a multiplet \Domult{} of highest weight modules.
We say that a set $\cm$ of highest weight modules forms a
{\it multiplet} if 1) $V\in\cm \Rightarrow \cm\supset \cm_V$,
where $\cm_V$ is the set of all highest weight modules
$V' \neq V$ such that either $V'$ is
isomorphic to a submodule of $V$ or $V$ is isomorphic to a
submodule of $V'$; 2) $\cm$ does not have proper subsets fulfilling 1). It is
convenient to depict a multiplet by a connected oriented graph, the
vertices of which correspond to the highest weight
modules and the arrows  between the vertices correspond to the embeddings between the modules.
Most often only the embeddings which are not
compositions of other embeddings are depicted, since these contain all the relevant information.

Further we shall restrict the notion of a multiplet to the
category of Verma modules. We also need the notion of ~{\it types of multiplets}.
We say that two multiplets belong to the same type, if they are depicted by the same
graph and differ only by the values of some parameter(s).

The classification can be summarized as follows. There are four types of
multiplets of reducible Verma modules: ~type ~$\cn$~ with four subtypes:
~$\cn_k$, $k=1,2,3,4$; ~type ~$\cf_{m_1,m_2}\,$, ($m_1,m_2\in\bbn$);
~type ~$\cs_{m_1,m_3}\,$, ($m_1,m_3\in\bbn$, ~$\ha (m_1 + m_3) \notin\bbz$);
~type ~$\cl$~ with two subtypes: ~$\cl_k\,$, $k=1,2$, as given explicitly below.

Multiplets of type ~$\cn$~  are given as follows. Fix $k=1,2,3,4$
to fix a subtype ~$\cn_k\,$.~ Then the multiplets of this subtype
are parametrized by the natural number $m_k$ and are given as follows:
\eqna\typa
$$\eqalignno{
V^{\L_k} ~&\longrightarrow~   V^{\L_k - m_k\a_k} \ ,
\quad  m_k(\L_k)= m_k\in\bbn \ , \quad m_j(\L_k)\notin\bbn, ~~j\neq k\ , &\typa a\cr
L_{\L_k} ~&=~ V^{\L_k} / V^{\L_k - m_k\a_k} \ .&\typa b\cr}$$
Note that we are using the convention that the arrows point to the embedded modules.
The modules $V^{\L_k - m_k\a_k}$ are irreducible. [The last statement is trivial for $k=1,2$
from \chan{}. For ~$k=3$~ we need to check that ~$m_j(\L_3)\notin-\bbn$ ~for~ $j=2,4$
which is easy since supposing that ~$m_2(\L_3)\in-\bbn$~ leads to ~$m_4(\L_3)\in\bbn$~
which is already excluded, and similarly interchanging $2\leftrightarrow 4$.
For ~$k=4$~ we need to check that ~$m_j(\L_4)\notin-\bbn$ ~for~ $j=1,3$ which is
proved similarly to the case $k=3$.]

\nt {\it Remark:}~ Under the same conditions  the ER counterpart
of \typa{} is given by \intt{}.~\dia

\medskip

Multiplets of type ~$\cf_{m_1,m_2}$~ are parametrized by two
natural numbers ~$m_1,m_2\,$. They are given as follows:
\fig{}{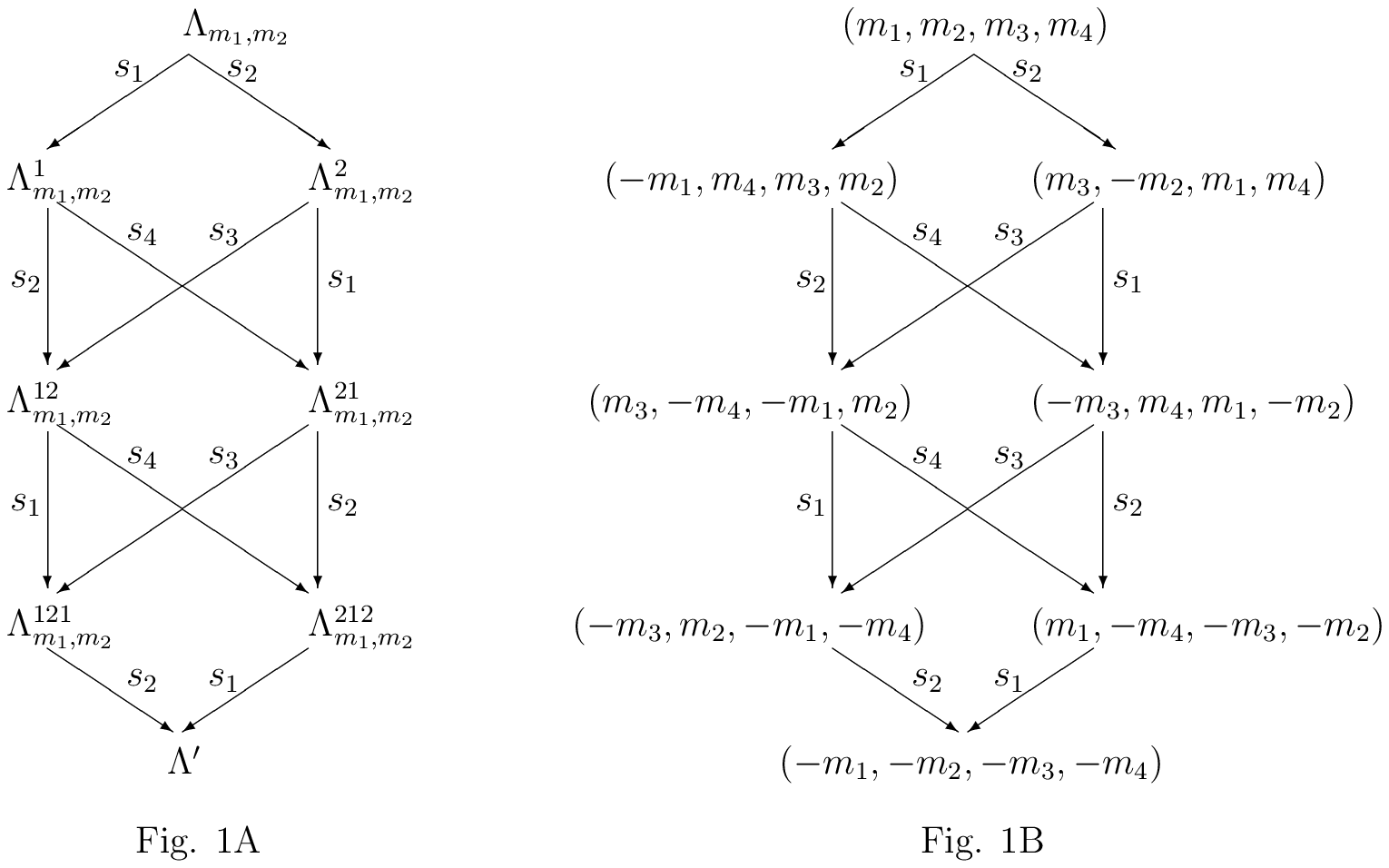}{12cm} \nt where we have given the multiplets
in two ways: on Fig.~1A the Verma modules are depicted by their
highest weights, while on Fig.~1B they are given by the explicit
signatures. Thus, we see that the parametrizing numbers $m_1,m_2$
are related to the Verma module $V^{\L_{m_1,m_2}}$ on the top:
~$m_k = m_k(\L_{m_1,m_2})$. The numbers at the embedding arrows
indicate w.r.t. which root is the embedding. All Verma modules of
these multiplets, except $V^{\L'}$, are reducible and their
weights are given explicitly as follows: \eqn\fdir{\eqalign{ &
\L_{m_1,m_2} \ , \quad m_k = m_k(\L_{m_1,m_2}) \in \bbn, ~~\forall
k \ , \cr & \L_{m_1,m_2}^i = \L_{m_1,m_2} - m_i\a_i\ , ~~~i=1,2 \
,\cr & \L_{m_1,m_2}^{ij} = \L_{m_1,m_2} - m_i\a_i - m_{j+2}\a_j \
, ~~~(i,j)=(1,2),(2,1)\ , \cr & \L_{m_1,m_2}^{iji} = \L_{m_1,m_2}
- (m_i+m_{i+2})\a_i - m_{j+2}\a_j \ , ~~~(i,j)=(1,2),(2,1)\  \cr}}
The weights of the irreducible modules $V^{\L'}$ are: ~$\L' =
\L_{m_1,m_2} - (m_1+m_3)\a_1 - (m_2+m_4)\a_2 = \L - 2m_4\a_1 -
m_3\a_2\,$. Note that only embeddings which are not compositions
of other embeddings are given on Fig. 1. In fact, several
composite embeddings according to the second lines of \svsv{c'}
and \svsv{d'} are present on Fig. 1.

\nt {\it Remark:}~~ To obtain the ER counterpart of Fig. 1 we need
to exclude four spaces with unphysical signatures
($m'_1\notin\bbn$), and to restore some operators which were
compositions on Fig. 1. Thus, the ER counterpart is given by
\diagr{}.~\dia

Multiplets of type ~$\cs_{m_1,m_3}$~ are parametrized by two natural numbers ~$m_1,m_3$~ of
different parity so that ~$\ha (m_1 + m_3) \notin\bbz$. They are given in Fig. 2
where as above we have given the multiplet in two ways,
and again the parametrizing numbers $m_1,m_3$ are related to the Verma module $V^{\L^s}$
on the top: ~$m_k = m_k(\L^s)$, $k=1,3$. Note that the supplementary condition on these numbers
ensure that ~$m_k = m_k(\L^s)\notin\bbn$, $k=2,4$, since ~$m_2(\L^s) = \ha (m_3-m_1)$,
~$m_4(\L^s) = \ha (m_3+m_1)$. The Verma modules of these multiplets, except $V^{\L''}$, are reducible
and their weights are given explicitly as follows:
\eqn\fdis{\eqalign{
&\L^s \ , \quad m_k = m_k(\L^s) \in \bbn, ~~k=1,3,  ~~~m_k = m_k(\L^s)\notin\bbn, ~~k=2,4\ , \cr
& \L^s_k = \L^s - m_k\a_k\ , ~~~k=1,3 \ .\cr}}
The weights of the irreducible modules $V^{\L''}$ are:
~$\L'' = \L^s - m_1\a_1 - m_3 \a_3\,$.
\fig{}{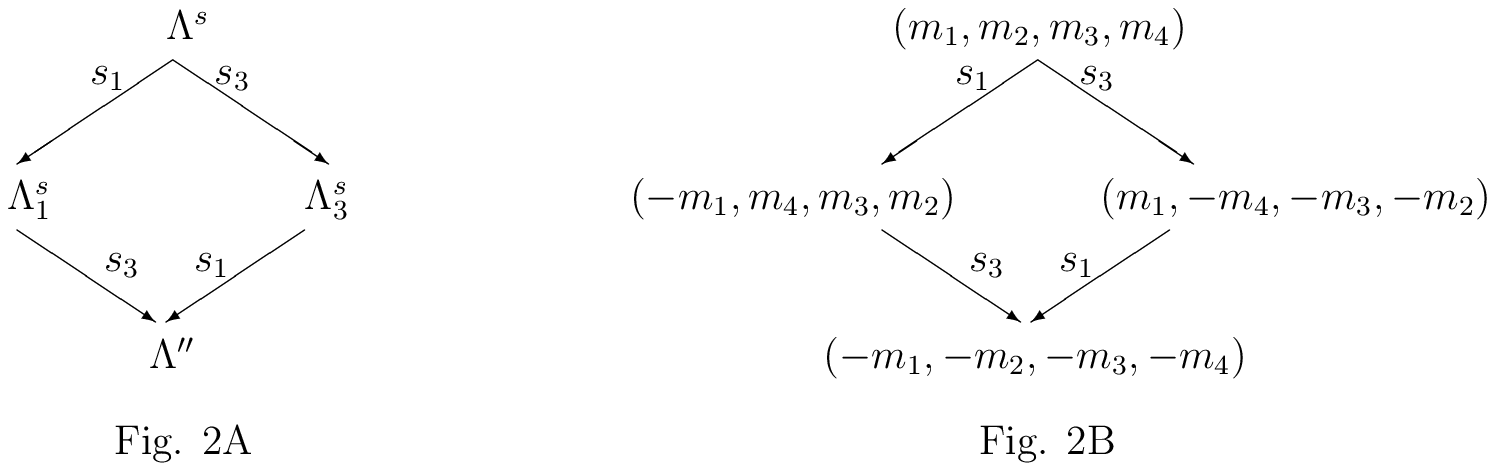}{11cm}

\nt {\it Remark:}~~ To obtain the ER counterpart of Fig. 2 we need
to exclude two spaces with unphysical signatures
($m'_1\notin\bbn$). Thus, the ER counterpart is given by \intt{}
for $\a=\a_3\,$.~\dia

Multiplets of type ~$\cl$~  are given as follows. Fix ~$k=1,2$~
to fix a subtype ~$\cl_k\,$.~ Then the multiplets
are parametrized by the natural number $m$ and are given as follows.

The multiplets of subtype ~$\cl_1\,$~ are given on Fig.3.
The weights of the reducible Verma modules are given as follows:
\eqn\reduc{\eqalign{
&\L^1_m \ , \quad m_2(\L^1_m) =m_4(\L^1_m) = m\in\bbn, ~~m_3(\L^1_m) = 2m, ~~m_1(\L^1_m)=0 \ , \cr
&\L^{12}_m = \L^1_m - m\a_2\ , \cr
&\L^{121}_m = \L^1_m - m\a_4\ . \cr}}
The weights of the irreducible modules $V^{\L'_m}$ are:
~~~$\L'_m = \L^1_m - 2m\a_3\,$. Note that this embedding picture is actually a special
case  of the factorization of the singular vector $v^{\a_3,m_3}$
according to the second line of \svsv{c'} for $m_1=0$, $m_2=m$.
\fig{}{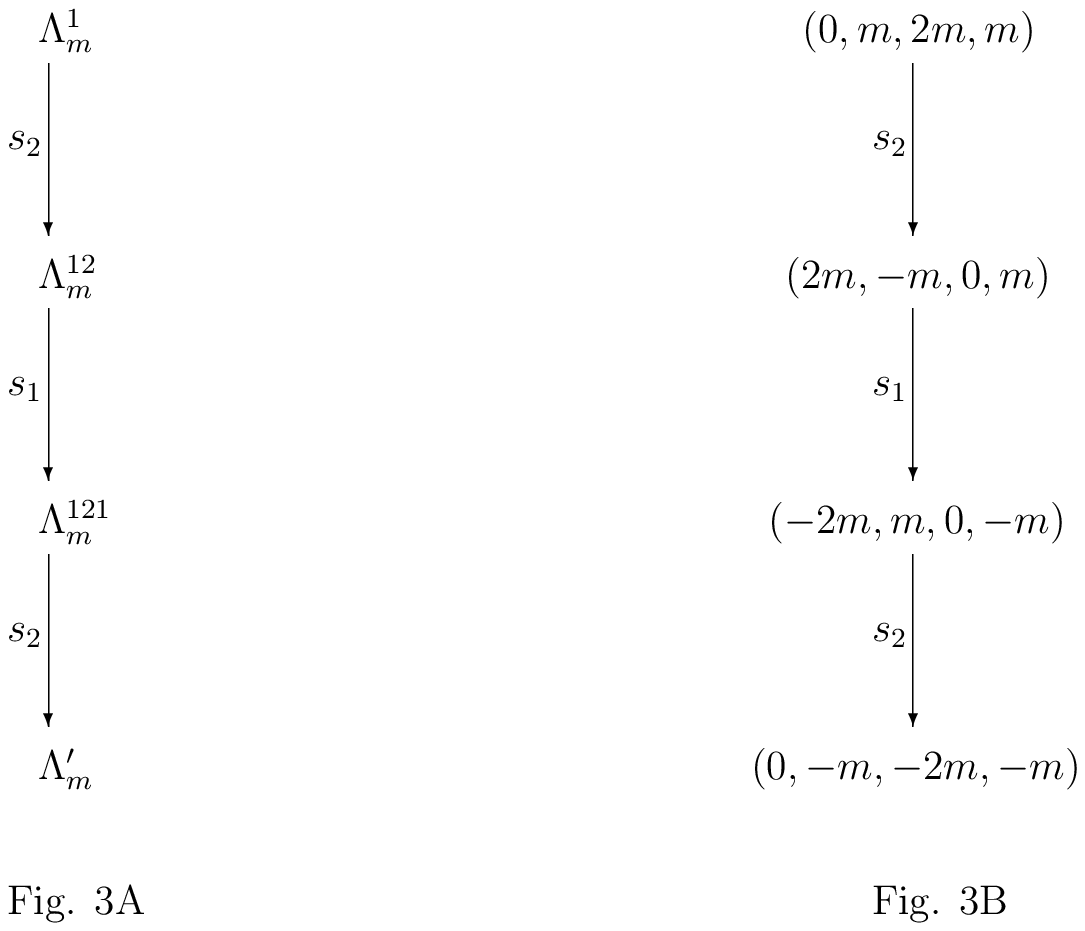}{8cm}

The multiplets of subtype ~$\cl_2\,$~ are given on Fig.4.
The weights of the reducible Verma modules are given as follows:
\eqn\reduca{\eqalign{
&\L^2_m \ , \quad m_1(\L^2_m) =m_3(\L^2_m) =  m_4(\L^2_m)=m\in\bbn, ~~m_2(\L^2_m)=0 \ , \cr
&\L^{21}_m = \L^2_m - m\a_1\ , \cr
&\L^{212}_m = \L^2_m - m\a_3\ . \cr}}
The weights of the irreducible modules $V^{\L''_m}$ are:
~~~$\L''_m = \L^2_m - m\a_4\,$. Note that this embedding picture is a special
case  of the factorization of the singular vector $v^{\a_4,m_4}$
according to the second line of \svsv{d'} for $m_2=0$, $m_1=m$.
\fig{}{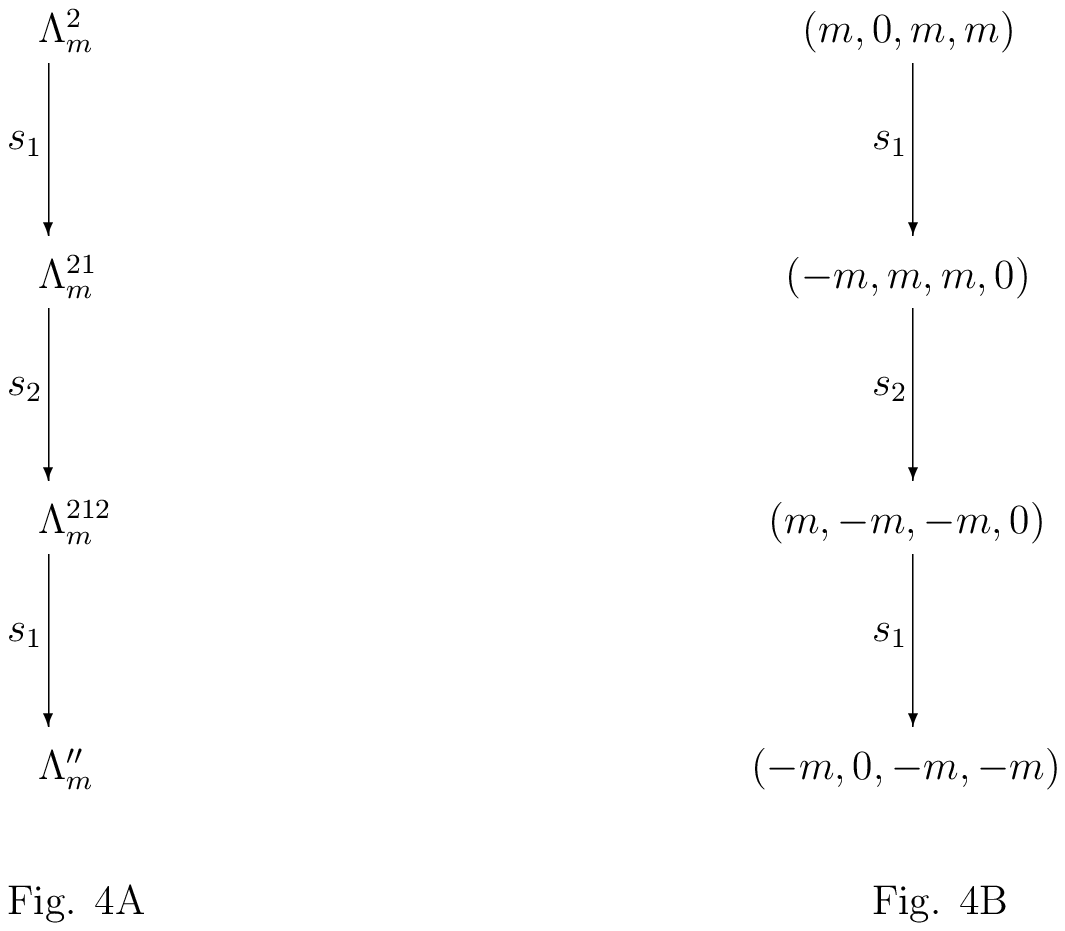}{8cm}

We can summarize the classification of the reducible Verma modules as follows:\nl
{\bf Proposition:}~~ All reducible Verma modules over ~$so(5,\bbc)$~
are explicitly parametrized by the
following highest weights: ~$\L_k$~ from \typa{},
$\L_{m_1,m_2}\,$, $\L_{m_1,m_2}^i\,$,
$\L_{m_1,m_2}^{ij}\,$,    $\L_{m_1,m_2}^{iji}\,$ from \fdir{},
$\L^s$,  $\L^s_k\,$ from \fdis{},
$\L^1_m\,$,  $\L^{12}_m\,$,  $\L^{121}_m\,$ from \reduc{},
$\L^2_m\,$,  $\L^{21}_m\,$,  $\L^{212}_m\,$ from \reduca{}.
The same highest weights parametrize all irreps $L_\L$ for which  $V^\L$
are reducible. All irreps are infinite-dimensional, except
~$L_{m_1,m_2} ~\equiv~ L_{\L_{m_1,m_2}}\,$. The latter are
the holomorphic finite-dimensional irreps of ~$so(5,\bbc)$~
of dimension ~$m_1m_2m_3m_4/6$, which are isomorphic to the
finite-dimensional non-unitary irreps of ~$so(3,2)$.~\dia

We note important property of reduction of multiplets. The multiplet ~$\cl_1\,$,
would be obtained from the multiplets of type ~$\cf_{m_1,m_2}$~ if we formally set ~$m_1=0$,
(and  then ~$m_2=m$), since the eight modules coincide pairwise reducing the multiplet
to four modules (the notation was chosen suggestively). Analogously, the multiplet ~$\cl_2\,$,
would be obtained from the multiplets of type ~$\cf_{m_1,m_2}$~ if we formally set ~$m_2=0$,
(and  then ~$m_1=m$). The opposite process would go from Fig. 2 if we suppose that $m_2,m_4\in\bbn$,
since then the two singular vectors $v^{\a_3,m_3}$ would decompose according to the second line
of \svsv{c'} and four new reducible modules would appear restoring Fig. 1.

\np

\newsec{Character formulae of AdS irreps and possible applications to integrability}

\newsubsec{Character formulae of AdS irreps.}~~~
The question of reducibility of Verma modules is closely related to the corresponding Weyl groups.
In particular, whenever \red{} is fulfilled, then the shifted weight is given by the corresponding Weyl reflection.
We recall that by definition the Weyl reflection in ~$\ch^*$~ is the following action:
\eqn\weyl{ s_\a (\L) ~\doteq~ \L - (\L, \a^\vee)\,\a  }
Using this we can define also a 'dot'-action:
\eqn\weyld{ s_\a \cdot (\L) ~\doteq~ s_\b\,(\L+\r) - \r \ .  }
Thus, we see that the shifted weight $\L'$ in \sing{} is given by (using \red):
\eqn\weyr{ \L' ~=~ \L -m\b ~=~ \L - (\L+\r, \b^\vee)\,\b ~=~ s_\b\, (\L+\r) -\r ~=~ s_\b\,\cdot\, (\L) }
The group generated by the Weyl reflections is called the Weyl group $W$. In fact, it is generated by
the simple reflections ~$s_{\a}\,$, where $\a$ runs through all simple roots.
Since the Weyl group is generated by the simple reflections then every element
~$w \in W$~ may be written as the product of some simple reflections. Every
such product which uses a minimal number of simple reflections is called a
reduced expression or reduced form for ~$w$~. The number of simple reflections
in a reduced form is called the length of ~$w$~ and is denoted by
~$\ell(w)$. In our case, ~$\cg^\bac ~=~ so(5,\bbc) ~=~ B_2\,$,
the Weyl group ~$W_{B_2}$~ has eight elements explicitly given in reduced form
by ($s_k \equiv s_{\a_k}$)~:
\eqn\weylg{
\eqalign{W_{B_2} ~=~ \{\
&1, ~s_1\,, ~s_2\,, ~s_1\,s_2\,, ~s_2\,s_1\,, ~s_3\,=\, s_2\,s_1\,s_2\,, ~
~s_4\,=\, s_1\,s_2\,s_1\,, \cr & s_1\,s_2\,s_1\,s_2 ~=~ s_2\,s_1\,s_2\,s_1 \ \}\ . }}

Let ~$\cg$~ be any simple Lie algebra.
Let ~$\G$~, (resp. ~$\G_+$~), be the set of all integral, (resp.
integral dominant), elements of ~$\ch^*$~, i.e., ~$\L
\in \ch^*$~ such that ~$(\L , \a_i^\vee) \in \bbz$~, (resp. ~$\bbz_+$~),
for all simple roots ~$\a_i$~. We recall that for each invariant
subspace ~$V\subset U(\cg_+)\otimes v_0 \cong V^\L$~ we have the
following decomposition
\eqn\decom{V ~=~ \mathop{\oplus}\limits_{\m\in\G_+} V_\m ~~,
~~~V_\m ~=~ \{ u \in V ~\vr ~H_ku = (\L + \m)(H_k)u, ~~
\forall ~H_k \} ~. }
(Note that ~$V_0 ~=~ \bbc ~v_0$~.)
Following \Dix,\Kac{} let ~$E(\ch^*)$~ be
the associative abelian algebra consisting of the series ~$\sum_{\m \in \ch^*}
c_{\m} \,e^\m$~, where ~$c_{\m} \in \bbc , ~c_{\m} = 0$~ for ~$\m$~ outside
the union of a finite number of sets of the form ~$D(\L) = \{ \m \in \ch^*
\vr \m \leq \L \}$~, using any ordering of ~$\ch^*\,$.

The character of ~$V$~ is defined by~:
\eqn\chav{ ch\ V ~~=~~ \sum_{\m \in\G_+} ({\rm dim}~V_\m)\,e^{\L+\m}
 ~~=~~ e^\L \sum_{\m\in\G_+} ({\rm dim}~V_\m)\,e^\m ~.}
We recall \Dix{} that for ~$V = V^\L$~ we have ~dim~$V_\m = P(\m)$~, ~$P(\m)$~ is
a generalized partition function, ~$P(\m) = \#$~ of ways ~$\m$~ can be
presented as a sum of positive roots ~$\b_j$~, each root taken with its
multiplicity ~$m_j = \dim ~\cg_{\b_j}$~, (here ~$m_j =1$~), ~$P(0)\equiv 1$~.
Analogously we use \Dix{} to obtain~:
\eqn\chave{ ch\ V^\L ~~=~~ e^\L\sum_{\m\in\G_+} P(\m)\, e^\m ~~=~~ e^\L
\prod_{\a \in\D^+}(1 - e^\a )^{-1} ~. }

The Weyl character formula for the finite-dimensional irreducible highest weight
representations over ~$\cg$~ has the form \Dix:
\eqn\chafd{ch\ L_\L ~~=~~ ch\ V^\L \sum_{w\in W}(-1)^{\ell(w)} \,e^{w\cdot\L - \L} ~~=~~
\sum_{w\in W}(-1)^{\ell(w)} ~ch\ V^{w\cdot\L} ~. }

Let ~$\cg = so(5,\bbc)$, or $\cg=so(3,2)$~.  Then the character formula
 for the finite-dimensional irreps \chafd{} can
be rewritten  as
\eqn\chafdn{\eqalign{ch\ L_{m_1,m_2} ~~=&~~ ch\ V^{\L_{m_1,m_2}}\ ( ~1 ~-~ e^{\a_1m_1} ~-~
e^{\a_2m_2} ~+~
e^{\a_1m_1}e^{\a_2(m_1+m_2)} ~+\cr &+~ e^{\a_1m_1+2m_2}e^{\a_2m_2}
~-~ e^{\a_4m_4}  ~-~ e^{\a_3m_3} ~+~
e^{2\a_1m_4}e^{\a_2m_3} ~) ~. }}
{\it Derivation of \chafdn:}~~
It is easy to derive the above formula using the embedding diagram on Fig. 1A.
We use the general fact that from \irrep{} follows:
\eqn\charg{ ch\ L_\L = ch\ V^\L - ch\ I^\L\ .}
Now it is clear that ~$ch\ I^{\L_{m_1,m_2}}$~ contains the terms
\eqn\subs{ ch\ V^{\L^1_{m_1,m_2}} + ch\ V^{\L^2_{m_1,m_2}}  }
since ~$V^{\L^1_{m_1,m_2}}$~ and ~$V^{\L^2_{m_1,m_2}}$~ are
submodules of ~$V^{\L_{m_1,m_2}}$, however,
the sum in \subs{} is larger than $ch\ I^{\L_{m_1,m_2}}$ because
of the overlap. Indeed, both ~$V^{\L^1_{m_1,m_2}}$~ and ~$V^{\L^2_{m_1,m_2}}$~
contain as submodules ~$V^{\L^{12}_{m_1,m_2}}$~ and ~$V^{\L^{21}_{m_1,m_2}}$.
Thus the latter are contained in \subs{} two times.
To correct this over-counting we should
subtract their characters and replace \subs{} by:
\eqn\subsa{ ch\ V^{\L^1_{m_1,m_2}}\ +\ ch\ V^{\L^2_{m_1,m_2}}
\ -\ ch\ V^{\L^{12}_{m_1,m_2}}\ -\ ch\ V^{\L^{21}_{m_1,m_2}} }
However, now there is under-counting since both
~$V^{\L^{12}_{m_1,m_2}}$~ and ~$V^{\L^{21}_{m_1,m_2}}$~
contain as submodules ~$V^{\L^{121}_{m_1,m_2}}$~ and ~$V^{\L^{212}_{m_1,m_2}}$.
 To correct this under-counting we should
add their characters and replace \subsa{} by:
\eqn\subsb{ ch\ V^{\L^1_{m_1,m_2}}\ +\ ch\ V^{\L^2_{m_1,m_2}}
\ -\ ch\ V^{\L^{12}_{m_1,m_2}}\ -\ ch\ V^{\L^{21}_{m_1,m_2}}
\ +\ ch\ V^{\L^{121}_{m_1,m_2}}\ +\ ch\ V^{\L^{212}_{m_1,m_2}}}
Now the module $V^{\L'}$ is over-counted. Thus, the final formula
for ~$ch\ I^{\L_{m_1,m_2}}$~ is:
\eqn\subm{\eqalign{ ch\ I^{\L_{m_1,m_2}} ~=&~ ch\ V^{\L^1_{m_1,m_2}}\ +\ ch\ V^{\L^2_{m_1,m_2}}
\ -\ ch\ V^{\L^{12}_{m_1,m_2}}\ -\ ch\ V^{\L^{21}_{m_1,m_2}} \ +\cr
&+\ ch\ V^{\L^{121}_{m_1,m_2}}\ +\ ch\ V^{\L^{212}_{m_1,m_2}}\ -\ ch\ V^{\L'} }}
Substituting \subm{} in \charg{}, using the explicit values of the various highest
weights from \fdir{}  we recover \chafdn{}.~\dia

The character formulae for the infinite-dimensional irreducible highest weight
representations   involve less terms than in \chafdn{} since the
maximal invariant submodules ~$I^\L$~ of ~$V^\L$~ are smaller. It is easy to
derive  these using the same considerations as above, so we just list the results.

\item{\bu} The character formulae for the irreps with highest weights
~$\L_1$~ from \typa{}, ~$\L^{212}_{m_1,m_2}$~ from \fdir{}, ~$\L^s_3$~
from \fdis{}, ~$\L^{12}_{m}$~ from \reduc{},
~$\L^{2}_{m}$~ and  ~$\L^{212}_{m}$~ from \reduca{}, are:
\eqn\redone{\eqalign{ ch\ L_\L ~=&~
\sum_{w\in W_1}(-1)^{\ell(w)} ~ch\ V^{w\cdot\L} ~=~ ch\ V^\L\ (1 - e^{m_1\a_1}) \ , \cr
&W_1 ~=~ \{\ 1,\ s_1\ \} }}
where ~$\L$~ denotes all highest weights under consideration and ~$m_1$~ should be replaced
by ~$m$~ for the cases from \reduc{} and \reduca{}.

\item{\bu} The character formula for the irreps with highest weights
~$\L_2$~ from \typa{}, ~$\L^{121}_{m_1,m_2}$~ from \fdir{},
~$\L^{1}_{m}$~ and  ~$\L^{121}_{m}$~ from \reduc{},
 ~$\L^{21}_{m}$~ from \reduca{},  is:
\eqn\redtwo{\eqalign{ ch\ L_\L ~=&~
\sum_{w\in W_2}(-1)^{\ell(w)} ~ch\ V^{w\cdot\L} ~=~ ch\ V^\L\ (1 - e^{m_2\a_2}) \ , \cr
&W_2 ~=~ \{\ 1,\ s_2\ \} }}
where ~$\L$~ denotes all highest weights under consideration and ~$m_2$~ should be replaced
by ~$m$~ for the cases from \reduc{} and \reduca{}.

\item{\bu} The character formula for the irreps with highest weights
~$\L_3$~ from \typa{},    ~$\L^s_1$~ from \fdis{}, is
\eqn\redthr{\eqalign{ ch\ L_\L ~=&~
\sum_{w\in W_3}(-1)^{\ell(w)} ~ch\ V^{w\cdot\L} ~=~ ch\ V^\L\ (1 - e^{m_3\a_3}) \ , \cr
&W_3 ~=~ \{\ 1,\ s_3\ \} }}
where ~$\L$~ denotes all highest weights under consideration.

\item{\bu} The character formula for the irreps with highest weights
~$\L_4$~ from \typa{}~  is
\eqn\redfou{\eqalign{ ch\ L_{\L_4} ~=&~
\sum_{w\in W_4}(-1)^{\ell(w)} ~ch\ V^{w\cdot\L_4} ~=~ ch\ V^{\L_4}\ (1 - e^{m_4\a_4}) \ , \cr
&W_4 ~=~ \{\ 1,\ s_4\ \} \ .}}

\item{\bu} The character formula for the irreps with highest weights
~$\L^s$~ from \fdis{}~  is
\eqn\redsin{\eqalign{ ch\ L_{\L^s} ~=&~
\sum_{w\in W^s}(-1)^{\ell(w)} ~ch\ V^{w\cdot\L^s} ~=\cr
=&~ ch\ V^{\L^s} ~( ~1 ~-~ e^{\a_1m_1} ~-~ e^{\a_3m_3} ~+~
e^{\a_1m_1+\a_3m_3}\ )\ , \cr
&W^s ~=~ \{\ 1 , ~s_1\,, ~s_3\,, ~s_1s_3\ \}  ~=~ W_1 \ \times\ W_3 }}

\item{\bu} The character formula for the irreps with highest weights
~$\L^{12}_{m_1,m_2}$~ from \fdir{}~  is
\eqn\redot{\eqalign{ ch\ L_{\L^{12}_{m_1,m_2}} ~=&~
\sum_{w\in W^{12}}(-1)^{\ell(w)} ~ch\ V^{w\cdot\L^{12}_{m_1,m_2}} ~=\cr
=&~ ch\ V^{\L^{12}_{m_1,m_2}} ~( ~1 ~-~ e^{\a_1m_3} ~-~ e^{\a_4m_2} ~+~
e^{\a_1m_3+\a_2m_2}\ )\ , \cr
&W^{12} ~=~ \{\ 1 , ~s_1\,, ~s_2s_1\,, ~s_1s_2s_1\ \} }}

\item{\bu} The character formula for the irreps with highest weights
~$\L^{21}_{m_1,m_2}$~ from \fdir{}~  is
\eqn\redto{\eqalign{ ch\ L_{\L^{21}_{m_1,m_2}} ~=&~
\sum_{w\in W^{21}}(-1)^{\ell(w)} ~ch\ V^{w\cdot\L^{21}_{m_1,m_2}} ~=\cr
=&~ ch\ V^{\L^{21}_{m_1,m_2}} ~( ~1 ~-~ e^{\a_2m_4} ~-~ e^{\a_3m_1} ~+~
e^{\a_1m_1+\a_2m_4}\ )\ , \cr
&W^{21} ~=~ \{\ 1 , ~s_2\,, ~s_1s_2\,, ~s_2s_1s_2\ \} }}

\item{\bu} The character formula for the irreps with highest weights
~$\L^{1}_{m_1,m_2}$~ from \fdir{}~  is
\eqn\redon{\eqalign{ ch\ L_{\L^{1}_{m_1,m_2}} ~=&~
\sum_{w\in W^{1}}(-1)^{\ell(w)} ~ch\ V^{w\cdot\L^{1}_{m_1,m_2}} ~=\cr
=&~ ch\ V^{\L^{1}_{m_1,m_2}} ~( ~1 ~-~ e^{\a_2m_4}
~-~ e^{\a_4m_2} ~+~ e^{\a_1m_3+\a_2m_4} ~+\cr &+~ e^{\a_2m_4+\a_4m_2} ~-~
e^{\a_1m_3+\a_2m_3}  \ )\ , \cr
&W^{1} ~=~ \{\ 1 , ~s_2\,, ~s_1s_2\,, ~s_2s_1s_2\,, ~s_1s_2s_1\,, ~s_2s_1s_2s_1
\ \} }}

\item{\bu} The character formula for the irreps with highest weights
~$\L^{2}_{m_1,m_2}$~ from \fdir{}~  is
\eqn\redtw{\eqalign{ ch\ L_{\L^{2}_{m_1,m_2}} ~=&~
\sum_{w\in W^{2}}(-1)^{\ell(w)} ~ch\ V^{w\cdot\L^{2}_{m_1,m_2}} ~=\cr
=&~ ch\ V^{\L^{2}_{m_1,m_2}} ~( ~1 ~-~ e^{\a_1m_3}
~-~ e^{\a_3m_1} ~+~ e^{\a_1m_3+\a_2m_4} ~+\cr &+~ e^{\a_1m_3+\a_3m_1} ~-~
e^{2\a_1m_4+\a_2m_4}  \ )\ , \cr
&W^{2} ~=~ \{\ 1 , ~s_1\,, ~s_2s_1\,, ~s_2s_1s_2\,, ~s_1s_2s_1\,, ~s_2s_1s_2s_1
\ \} }}

Note that each of the Weyl groups ~$W_k\,$, $k=1,2,3,4$, is isomorphic to
the ~$A_1\, =\, sl(2)$~ Weyl group, while the Weyl group ~$W^s$~ is  the direct product of
two such ~$A_1$~ Weyl groups. In contrast,
the subsets of ~$W$~ over which is carried the summation in the
last four cases, namely, ~$W^{1}$, $W^{2}$, $W^{12}$, $W^{21}$~ are not considered
subgroups of ~$W$, ~since then the
elements of these subsets will generate the whole ~$W$.

\newsubsec{Character formulae of positive energy UIRs.}~~~
Here we apply the character formulae of the previous subsection to the
positive energy UIRs of ~$so(3,2)$.

\newsubsubsec{Rac.} ~~  We have ~$m_1 ~=~ 1$~, ~$m_2 ~=~ 1/2$, i.e., we have a special case of
\redsin{}:
\eqna\charac
$$\eqalignno{ch~L_{{\rm Rac}} ~~&=~~ ch~V^{\L^s}
~( ~1 ~-~ e^{\a_1} ~-~ e^{2\a_3} ~+~ e^{\a_1+2\a_3}\ )
  ~= &\charac a\cr
&=~~ e(\L^s) ~( ~1 ~-~ e^{2\a_3} ~) / (1 - e^{\a_2}) (1 - e^{\a_3}) (1 - e^{\a_4}) ~=
&\charac b\cr
&=~~ e(\L^s) ~( ~1 ~+~ e^{\a_3} ~) / (1 - e^{\a_2}) (1 - e^{\a_4})  ~=
&\charac c\cr
&=~~ e(\L^s) ~\sum_{n=0}^{\infty} ~ e^{n\a_3} ~\sum_{p=-n}^n ~e^{p\a_1}   ~=
&\charac d\cr
&=~~ e(\L) ~\sum_{n=0}^{\infty} ~\sum_{p=-n}^n ~ e^{(n-\vr p\vr)\a_3}
~t'^{\vr p\vr} ~, &\charac e\cr}$$
where
$$t' ~~=~~ \cases{  e(\a_4) &~~for ~~$p \geq 0$ ~, \cr
  e(\a_2) &~~for ~~$p < 0$ ~. \cr}$$
Character formula \charac{} is equivalent to the spectrum
description given in \DoSe{} and clearly each term has different weight from all others,
which explains the terminology of singleton.

\newsubsubsec{Di.} ~~ We have ~$m_1 ~=~ 2$~, ~$m_2 ~=~ -1/2$, i.e., again a special case of
\redsin{}:
\eqna\chadi
$$\eqalignno{ch~L_{{\rm Di}} ~~&=~~ ch~V^{\L^s}
~( ~1 ~-~ e^{2\a_1} ~-~ e^{\a_3} ~+~ e^{2\a_1+\a_3}\ )
  ~= &\chadi a\cr
&=~~ e(\L^s) ~( ~1 ~-~ e^{2\a_1} ~) / (1 - e^{\a_1})(1 - e^{\a_2})  (1 - e^{\a_4}) ~=
&\chadi b\cr
&=~~ e(\L^s) ~( ~1 ~+~ e^{\a_1} ~) / (1 - e^{\a_2})  (1 - e^{\a_4}) ~=
&\chadi c\cr
&=~~ e(\L^s) ~\sum_{n=0}^{\infty} ~ e^{n\a_3} ~\sum_{p=-n}^{n+1} ~e^{p\a_1}   ~=
&\chadi d\cr
&=~~ e(\L^s) ~\sum_{n=0}^{\infty} ~e^{n\a_2} ~\sum_{r=0}^{2n+1}~ e^{r\a_1} ~) ~.
&\chadi e\cr}$$
Character formula \chadi{} is equivalent to the singleton spectrum
description given in \DoSe{}.

\newsubsubsec{Spin zero.}~~
Next we consider the case ~$s_0 ~=~ 0$, ~$E_0 >1/2$, cf. \unita{}, \sv{} and the
text in-between. We have only the singular vector ~$v^{\a_1,1}$.
This is clear for ~$E_0 \neq 1$, while for ~$E_0 ~=~ 1$~ one should
note that for ~$s_0=0$~ the singular vectors ~$v^{\a_3,1}$~ in \sv{c} and ~$v^{\a_4,1}$~ in \sv{d}
are descendants of $v^{\a_1,1}$. In fact, for $E_0\neq 1$ the Verma module is ~$V^{\L_1}$~
from a multiplet of subtype ~$\cn_1$~ for parameter $m_1=1$, while for $E_0= 1$ the Verma module
is ~$V^{\L^2_1}$~
from a multiplet of  subtype ~$\cl_2$~ for parameter $m=1$. Thus, the character formula is \redone.

\newsubsubsec{Spin 1/2.}~~
Analogously for ~$s_0 ~=~ 1/2, ~E_0 > 1$~, we have only the singular vector ~$v^{\a_1,2}$.
 This is clear for ~$E_0 \neq 3/2$, while for ~$E_0 ~=~ 3/2$~ one
should note that for ~$s_0=1/2$~ the singular vector $v^{\a_4,1}$~ in \sv{d}
is  descendant of $v^{\a_1,2}$. In fact, for $E_0\neq 3/2$ the Verma module is ~$V^{\L_1}$~
from a multiplet of subtype ~$\cn_1$~ for parameter $m_1=2$,
while for $E_0= 3/2$ the Verma module is ~$V^{\L^{12}_1}$~
from a multiplet of  subtype ~$\cl_1$~ for parameter $m=1$.
 Thus, the character formula is \redone{} as in the previous case.

\newsubsubsec{Higher spins.}~~
Analogously for ~$s_0 \geq 1, ~E_0 > s_0+1$, cf. \unita{b},
we have only the singular vector ~$v^{\a_1,m_1}$, $m_1=2s_0+1$, since
~$m_2,m_3,m_4\notin\bbn$. Thus,  the Verma module is ~$V^{\L_1}$~
from a multiplet of subtype ~$\cn_1$~ for parameter $m_1\geq 3$,
and the character formula is \redone{} as in the previous case.

\newsubsubsec{Massless irreps.}~~
Finally we consider the massless representations with ~$E_0 ~=~ s_0 +1$~, ~$s_0
\geq 1$. We have two singular vectors: ~$v^{\a_1,m_1}$, $m_1=2s_0+1$,
and $v^{\a_4,1}$. The signature is: ~$(m_1,m_2,m_3,m_4) ~=~ (2s_0+1,-2s_0,1-2s_0,1)$.
This signature appears as the Verma module ~$V^{12}_{2s_0-1,1}$~ of the multiplet
~$\cf_{2s_0-1,1}\,$. Thus, the character formula (found first in \DoSe)
is a special case of \redot:
\eqn\redots{\eqalign{ ch\ L_{\L^{12}_{m-2,1}} ~=&~
  ch\ V^{\L^{12}_{m-2,1}} ~( ~1 ~-~ e^{m\a_1} ~-~ e^{\a_4} ~+~
e^{m\a_1+\a_2}\ )\ , \cr
& m = m_1 = 2s_0+1}}
Thus, the massless UIRs are in one-parameter family of multiplets ~$\cf_{2s_0-1,1}\,$~
where in the Verma modules ~$V^{\L_{2s_0-1,1}}$~ on the top of the multiplets
are found the finite-dimensional irreps of dimension: ~$s_0 (4s^2_0-1)/3$, ~$s_0=1,\trha,\ldots$
~For the lowest possible value  $s_0=1$ the finite-dimensional
irrep    is the trivial one-dimensional irrep
of $so(5,\bbc)$ (and of $so(3,2)$).

\newsubsec{Possible applications to integrability.}~~~
The classical and quantum integrability/solvability of
Calogero-Moser-Sutherland systems \refs{\Cala,\Sut,\Moser,\CMR,\Calb}
has attracted a lot of attention for the last 30 years.
A very important contribution for the group-theoretic understanding was made by
Olshanetsky and Perelomov who extended the original models from the A root system
to the B, C and D root systems \OPa,\OPb. For a recent overview we refer to \SaTa{}
and references therein.

For our setting the interesting aspect is the relation
between Coxeter groups and integrability of these systems.
More specifically, we are interested in
the relation of integrable systems to characters of irreps of
simple Lie algebras, cf. \FFP{} and references therein.
To illustrate the possible applications of the results of this paper
to integrability we first look at the case of Dirac singletons.

In both singleton cases the character formula is given in \redsin{} and
involves summation over the group ~$W^s$~ which is the direct product of
two ~$A_1$~ Weyl groups. The group ~$W^s$~ can be characterized as
generated by the short root reflections  ~$s_1$~ and ~$s_3\,$.
Actually, this is a manifestation of a general phenomena since the sets
of short and long roots are invariant
under the action of the whole Weyl group. Thus, we may also write:
\eqn\deco{  W_{B_2}^{\rm short} ~=~ W_{A_1} ~\times~ W_{A_1}   \ .}

Thus, irreps for which holds \redsin{}, in particular,
the two Dirac singletons, would be related to integrable systems described by the orbit of short roots
of the ~$B_2$~ root system \OPa,\OPb,\SaTa.

The irreps with character formulae \redone,\redtwo,\redthr,\redfou{}
would be related to integrable systems described by the ~$A_1$~ root system.

All the above could also related to different Toda-like models (for a recent review,
cf. \Corr)
through their character formulae similarly to the application of Virasoro characters in \ByFr,
which should be extendable to the supersymmetry setting \Doch.

More interesting would be models using character formulae
\redot,\redto,\redon,\redtw{} since in these cases summation is
over  subsets of ~$W_{B_2}$~ which can  not be considered
subgroups of ~$W_{B_2}\,$. Thus, these would be new models!

\vskip 10mm

\nt {\bf Acknowledgments.}\nl
The author would like to thank for hospitality
the Abdus Salam International Center for Theoretical Physics,
where part of the work was done.
This work was supported in part by
the Bulgarian National Council for Scientific Research, grant
F-1205/02,  the Alexander von Humboldt
Foundation in the framework of the
Clausthal-Leipzig-Sofia Cooperation,
the TMR Network EUCLID, contract HPRN-CT-2002-00325,
 and the European RTN  'Forces-Universe',
contract MRTN-CT-2004-005104.

\vskip 10mm

\np \parskip=0pt \listrefs
\nd